\documentclass[3p]{elsarticle}   
\usepackage{hyperref}

\usepackage{ucs}
\usepackage[utf8x]{inputenc}
\usepackage[T1]{fontenc}
\PrerenderUnicode{ÉéÀàÈè¼}

\usepackage{xspace}

\usepackage{amsmath}\usepackage{amsthm}\usepackage{amssymb}  
\usepackage{makeidx} 
\usepackage{amsgen} 
\usepackage{currfile} 
\usepackage{longtable}

\usepackage{fancyvrb}

\usepackage{pifont} 
\usepackage{tikz}  \usetikzlibrary{intersections}
\usetikzlibrary{decorations.pathreplacing}
\usetikzlibrary{calc}
\usetikzlibrary{arrows}
\tikzset{minimum size=0}

\graphicspath{{.}}

\usepackage{subcaption}

\usepackage{tabularx}
\usepackage{enumitem}
\usepackage[frozencache]{minted}

\usepackage{h_color} 
\usepackage{h_vocabulary} 
\usepackage{h_graphics} 
\usepackage{h_agc} 

\usepackage{Add}

\providecommand{\Latin}[1]{\text{#1}\xspace} 

\providecommand{\ie}{\Latin{i.e.}}

\newtheorem{theorem}{Theorem}

\newtheorem{lemma}[theorem]{Lemma}

{\theoremstyle{definition}
}

{\theoremstyle{definition}
}

\newcommand{\RefFigure}[1]{Figure\,\ref{#1}}
\newcommand{\RefFig}[1]{Fig.\,\ref{#1}}

\newcommand{\RefLem}[1]{Lem.\,\ref{#1}}

\newcommand{\RefSection}[1]{Section\,\ref{#1}}
\newcommand{\RefSec}[1]{Sect.\,\ref{#1}}

\newcommand{\RefAlgorithm}[1]{Algorithm\,\ref{#1}}
\newcommand{\RefAlgo}[1]{Algo.\,\ref{#1}}

\newenvironment{MSlist}[1][]{\begingroup\small\footnotesize\scriptsize\begin{tabular}{@{}r@{\,}c@{\,}>{\begin{math}}c<{\end{math}}@{}}&\,{\small\footnotesize\bf #1Meta-signal\,}&\,\text{\small\footnotesize\bf Speed}\!\\[.1em]
    }{\end{tabular}\endgroup}

\newenvironment{CRlist}{\begingroup\small\footnotesize\scriptsize\begin{tabular}{@{}r@{\,}>{\raggedleft\arraybackslash\{\,}r<{\,\}}@{\,$\to$\,\{\,}>{\raggedright\arraybackslash}l<{\,\}}}}{\end{tabular}\endgroup}

\newenvironment{AugCRlist}{\begingroup\small\footnotesize\scriptsize\begin{tabular}{@{}r@{\,}>{\raggedleft\arraybackslash\{\,}r<{\,\}}@{\,}r@{\,\{\,}>{\raggedright\arraybackslash}l<{\,\}}}}{\end{tabular}\endgroup}

\newcommand{\NTVS}{\vspace{-1mm}}

\providecommand{\SetUnitlength}[1]{\setlength{\unitlength}{#1}\ifx\tikzpicture\undefined\relax\else\tikzset{x=#1}\tikzset{y=#1}\fi\ifx\PSTricksLoaded\undefined\relax\else\psset{unit=#1}\fi}

\tikzstyle{ray_style}=[fill=yellow!60!white,draw=none]

\newcommand{\foreachASig}[1]{&\csname Sig#1\endcsname &0\\
  &\csname SigSL#1\endcsname &-1\\
  &\csname SigSR#1\endcsname &1
}

\providecommand{\mathxspace}[1]{\ensuremath{#1}\xspace}
\providecommand{\mathBBxspace}[1]{\mathxspace{\mathbb{#1}}}
\providecommand{\mathCalxspace}[1]{\mathxspace{\mathcal{#1}}}

\JDLvocabulary{\NaturalSet}{\mathBBxspace{N}}{}{Natural set}
\JDLvocabulary{\RationalSet}{\mathBBxspace{Q}}{}{Rational set}
\JDLvocabulary{\RealSet}{\mathBBxspace{R}}{}{Real set}

\JDLvocabulary{\SpeedSet}{\mathCalxspace{S}}{}{(finite) Set of speeds}

\JDLvocabulary{\SpaceCoordinate}{\mathxspace{x}}{}{Some spatial coordinate}
\JDLvocabulary{\TimeCoordinate}{\mathxspace{t}}{}{Some temporal coordinate}
\JDLvocabulary{\Duration}{\mathxspace{\Delta t}}{}{Some duration}

\JDLvocabulary{\CollisionDepth}{\JDLvocabularyMathXspace{d}}{}{the depth of a collision}

\JDLvocabulary{\Slope}{\JDLvocabularyMathXspace{\alpha}}{}{the targeted slope, used in the specification of the problem as well as in proofs}
\JDLvocabulary{\generalWidth}{\JDLvocabularyMathXspace{w}}{}{the maximum width of the general, used in the specification of the problem}

\JDLvocabulary{\InputS}{\JDLvocabularyMathXspace{l}}{}{Input parameter of the algorithm (left height)}
\JDLvocabulary{\InputT}{\JDLvocabularyMathXspace{r}}{}{Input parameter of the algorithm (right height)}
\JDLvocabulary{\Unit}{\mathxspace{u}}{}{the unit}
\JDLvocabulary{\InputSInitial}{\mathxspace{\InputS_0}}{}{an input parameter of the initial call of the algorithm (left height)}
\JDLvocabulary{\InputTInitial}{\mathxspace{\InputT_0}}{}{an input parameter of the initial call of the algorithm (right height)}

\JDLvocabulary{\RegisterS}{\mathxspace{x_s}}{}{an input parameter of the algorithm (left height)}
\JDLvocabulary{\RegisterT}{\mathxspace{x_t}}{}{an input parameter of the algorithm (right height)}
\JDLvocabulary{\RegisterU}{\mathxspace{x_u}}{}{the unit}

\JDLvocabulary{\RegisterM}{\mathxspace{t_m}}{}{a temporal register for intermediate computation}
\JDLvocabulary{\Register}{\mathxspace{x}}{}{a generic register, can be (store) \InputS, \InputT or \Unit}

\JDLvocabulary{\AGCaugmentedMetaSignalSet}{\JDLvocabularyMathXspace{M'}}{}{Augmented Signal machine meta-signal label set (the finite set ignoring parameters)}
\JDLvocabulary{\AGCaugmentedMetaSignal}{\JDLvocabularyMathXspace{\AGCmetaSignal_{\InputS}^{\InputT}}}{}{Some meta-signal}

\JDLvocabulary{\goto}{\mathxspace{goto}}{}{goto}
\JDLvocabulary{\iftxt}{\mathxspace{if}}{}{if}

\JDLvocabulary{\Up}{\mathxspace{Up}}{}{up}
\JDLvocabulary{\Split}{\mathxspace{Split}}{}{split}
\JDLvocabulary{\main}{\mathxspace{main}}{}{main}

\tikzAGCmakeSignal{Border}{border}{Black,thick}
\JDLvocabulary{\AGCconfigurationSMSymbol}{\ensuremath{a}}{}{Some configuration of \AGCmachine}

\JDLvocabulary{\AGCaugSignalDomainFun}{\JDLvocabularyMathXspace{D}}{}{Augmented Signal machine domain function}

\tikzAGCmakeSignal{ZZle}{\AGCRightWardSlow{le}}{}
\tikzAGCmakeSignal{ZZri}{\AGCLeftWardSlow{ri}}{}
\tikzAGCmakeSignal{Zig}{\AGCRightWard{zig}}{}
\tikzAGCmakeSignal{Zag}{\AGCLeftWard{zag}}{}

\JDLvocabulary{\StepDelay}{\JDLvocabularyTextXspace{{\rm\textsl{\texttt{delay}}}}}{}{Delay step of the augmented signal machine}
\JDLvocabulary{\StepSplit}{\JDLvocabularyTextXspace{{\rm\textsl{\texttt{split}}}}}{}{Split step of the augmented signal machine}

\newcommand{\AGCLeftSide}[1]{\AGCOverLine{#1}{densely dotted,right to-}}  
\newcommand{\AGCRightSide}[1]{\AGCOverLine{#1}{densely dotted,-left to}}
\newcommand{\AGCLeftRightSide}[1]{\AGCOverLine{#1}{densely dotted}}

\newcommand{\IndexSB}{\ensuremath{_{\text{sb}}}}
\newcommand{\IndexS}{\ensuremath{_{\text{s}}}}
\newcommand{\IndexSplit}{\ensuremath{_{\text{sp}}}}
\newcommand{\IndexDelay}{\ensuremath{_{\text{dl}}}}
\newcommand{\IndexOne}{\ensuremath{_1}}
\newcommand{\IndexBack}{\ensuremath{_{\text{back}}}}

\newcommand{\IndexSlow}{\ensuremath{_{\text{sl}}}}

\newcommand{\AugmentedPoliceModification}[1]{\textit{#1}}
\tikzAGCmakeSignal{ABounceRSlow}{\AGCRightWardSlow{\AugmentedPoliceModification{bounce\IndexSlow}}}{VeryDarkGreen,thick}
\tikzAGCmakeSignal{ABounceR}{\AGCRightWard{\AugmentedPoliceModification{bounce}}}{VeryDarkGreen,thick}
\tikzAGCmakeSignal{ABounceL}{\AGCLeftWard{\AugmentedPoliceModification{bounce}}}{Purple,thick}

\newcommand{\ParamLR}{\ensuremath{_{l}^{r}}}
\newcommand{\ParamLRmO}{\ensuremath{_{l{-}1}^{r{-}1}}}
\newcommand{\ParamLLR}{\ensuremath{_{l{+}l{-}1}^{r{+}l{-}1}}}
\newcommand{\ParamLRR}{\ensuremath{_{l{+}r{-}1}^{r{+}r{-}1}}}
\tikzAGCmakeSignal{Delay}{\AugmentedPoliceModification{\AugmentedPoliceModification{delay}\ParamLR}}{thick, densely dotted}
\tikzAGCmakeSignal{DelaymO}{\AugmentedPoliceModification{\AugmentedPoliceModification{delay}\ParamLRmO}}{thick, densely dotted}
\tikzAGCmakeSignal{SplitR}{\AGCRightWardSlow{\AugmentedPoliceModification{split}\ParamLR}}{blue,thick}
\tikzAGCmakeSignal{SplitL}{\AGCLeftWardSlow{\AugmentedPoliceModification{split}\ParamLR}}{red,thick}
\tikzAGCmakeSignal{SplitXL}{\AGCLeftWardSlow{\AugmentedPoliceModification{split}\ParamLLR}}{red,thick}
\tikzAGCmakeSignal{SplitXR}{\AGCRightWardSlow{\AugmentedPoliceModification{split}\ParamLRR}}{blue,thick}

\newcommand{\ExpBot}{\ensuremath{^{\text{bot}}}}
\newcommand{\ExpTop}{\ensuremath{^{\text{top}}}}

\tikzAGCmakeSignal{BounceRSlowTop}{\AGCRightWardSlow{bounce$^{\text{top}}\IndexSlow$}}{VeryDarkGreen,thick}\tikzAGCmakeSignal{BounceRSlowBot}{\AGCRightWardSlow{bounce$^{\text{bot}}\IndexSlow$}}{VeryDarkGreen,thick}\tikzAGCmakeSignal{BounceRTop}{\AGCRightWard{bounce\ExpTop}}{DarkGreen,thick}
\tikzAGCmakeSignal{BounceRBot}{\AGCRightWard{bounce\ExpBot}}{DarkGreen,thick}
\tikzAGCmakeSignal{BounceLTop}{\AGCLeftWard{bounce\ExpTop}}{DarkGreen,thick}
\tikzAGCmakeSignal{BounceLBot}{\AGCLeftWard{bounce\ExpBot}}{DarkGreen,thick}
\tikzAGCmakeSignal{BounceRSlowI}{\AGCRightWardSlow{bounce$^{i}\IndexSlow$}}{VeryDarkGreen,thick}
\tikzAGCmakeSignal{BounceRI}{\AGCRightWard{bounce$^{i}$}}{DarkGreen,thick}
\tikzAGCmakeSignal{BounceRJ}{\AGCRightWard{bounce$^{j}$}}{DarkGreen,thick}
\tikzAGCmakeSignal{BounceLI}{\AGCLeftWard{bounce$^{i}$}}{DarkGreen,thick}
\tikzAGCmakeSignal{BounceLJ}{\AGCLeftWard{bounce$^{j}$}}{DarkGreen,thick}
\JDLvocabulary{\SpeedBounce}{\JDLvocabularyTextXspace{{\AGCspeed}$_{\text{bounce}}$}}{}{absolute speed of bouncing signals}
\newcommand{\SpeedBouncevalue}{3}
\JDLvocabulary{\SpeedBounceSlow}{\JDLvocabularyTextXspace{{\AGCspeed}$^{\text{slow}}_{\text{bounce}}$}}{}{absolute speed of slow bouncing signals}
\newcommand{\SpeedBounceSlowvalue}{3/2}

\JDLvocabulary{\WidthTree}{\JDLvocabularyMathXspace{w_t}}{}{Width of the tree macro-signal}
\JDLvocabulary{\HeightBouncing}{\JDLvocabularyMathXspace{h_b}}{}{Height of the bouncing pairs of signals}
\JDLvocabulary{\NewWidthTree}{\JDLvocabularyMathXspace{w_{t'}}}{}{Width of the tree macro-signal after a step}
\JDLvocabulary{\NewHeightBouncing}{\JDLvocabularyMathXspace{h_{b'}}}{}{Height of the bouncing pairs of signals after a step}

\JDLvocabulary{\refraction}{\JDLvocabularyTextXspace{{\rm\textsl{\texttt{refraction}}}}}{}{a refraction}
\JDLvocabulary{\reflection}{\JDLvocabularyTextXspace{{\rm\textsl{\texttt{reflection}}}}}{}{a reflection}

\newcommand{\definMSinBand}[3]{\tikzAGCmakeSignal{#1}{#2}{#3}
  \tikzAGCmakeSignal{SL#1}{\AGCLeftSide{#2}}{#3}
  \tikzAGCmakeSignal{SR#1}{\AGCRightSide{#2}}{#3}
}

\JDLvocabulary{\SPEEDCOMPUTE}{\JDLvocabularyTextXspace{{\AGCspeed}$_{\text{test}}$}}{}{speed of test computation}
\newcommand{\SPEEDCOMPUTEvalue}{2}

\definMSinBand{Tree}{tree}{Blue,thick}\definMSinBand{TreeDelay}{tree\IndexDelay}{Blue,thick,dotted}\definMSinBand{One}{one}{DarkBrown,thick}\definMSinBand{Two}{two}{DarkOrange,thick}\definMSinBand{L}{l}{Blue,thick}\definMSinBand{R}{r}{Blue,thick}
\definMSinBand{Test}{test}{DeepPurple,thick}\definMSinBand{TestSPL}{test\IndexSplit}{DeepPurple, dashed,thick}\definMSinBand{TestDEL}{test\IndexDelay}{DeepPurple, dotted,thick}\definMSinBand{Bound}{bound}{VeryDarkGrey,thick}\definMSinBand{BoundSPL}{bound\IndexSplit}{VeryDarkGrey, dashed,thick}\definMSinBand{BoundDEL}{bound\IndexDelay}{VeryDarkGrey, dotted,thick}
\definMSinBand{PreComputeMove}{updt$^{*}\IndexDelay$}{DarkGreen,thick}
\definMSinBand{ComputeMove}{updt$^2\IndexDelay$}{DarkGreen,thick}\definMSinBand{ComputeMoveOne}{updt$^1\IndexDelay$}{Green,thick}\definMSinBand{OneBack}{one$_{\text{back}}$}{DarkBrown,thick}\definMSinBand{TwoBack}{two$_{\text{back}}$}{DarkOrange,thick}\definMSinBand{LBack}{l$_{\text{back}}$}{Blue,thick}\definMSinBand{RBack}{r$_{\text{back}}$}{Blue,thick}\definMSinBand{Minus}{minus$^2$}{DarkGreen,thick}\definMSinBand{MinusOne}{minus$^1$}{Green,thick}\definMSinBand{TestStart}{\JDLvocabularyTextXspace{test$_{\text{start}}$}}{Purple,thick}
\JDLvocabulary{\SpeedBounceFASTSHRINK}{\JDLvocabularyTextXspace{\AGCspeed\!\!$^{\text{fast}}_{\text{shrink}}$}}{}{speed of the intermediate \string\SigABounceRSlow{} signal when shrinking into delay}
\newcommand{\SpeedBounceFASTSHRINKvalue}{3/5}
\JDLvocabulary{\SPEEDSHRINK}{\JDLvocabularyTextXspace{\AGCspeed\!\!$_{\text{shrink}}$}}{}{speed of intermediate signals when shrinking into delay}
\newcommand{\SPEEDSHRINKvalue}{3/7} \JDLvocabulary{\SPEEDDSEPTWO}{\JDLvocabularyTextXspace{\AGCspeed\!\!$_{\text{shrink}}$}}{}{speed of dsep2}

\tikzAGCmakeSignal{WallDelay}{wall\IndexDelay}{densely dotted,thick}
\tikzAGCmakeSignal{BounceFSR}{\AGCRightWard{bounce\IndexS}}{VeryDarkGreen,thick} \tikzAGCmakeSignal{BoundDelaySR}{\AGCRightWardSlow{bound\IndexS\IndexDelay}}{VeryDarkGrey, densely dotted,thick}
\tikzAGCmakeSignal{BounceDelayL}{\AGCLeftWard{bounce\IndexDelay}}{VeryDarkGreen, densely dotted,thick} \tikzAGCmakeSignal{OneSR}{\AGCRightWardSlow{one\IndexS}}{DarkBrown,thick}
\tikzAGCmakeSignal{OneSL}{\AGCLeftWardSlow{one\IndexS}}{DarkBrown,thick}
\tikzAGCmakeSignal{TwoSR}{\AGCRightWardSlow{two\IndexS}}{DarkOrange,thick}
\tikzAGCmakeSignal{TwoSL}{\AGCLeftWardSlow{two\IndexS}}{DarkOrange,thick}
\tikzAGCmakeSignal{RSR}{\AGCRightWardSlow{{\SigR}\IndexS}}{blue,thick}
\tikzAGCmakeSignal{RSL}{\AGCLeftWardSlow{{\SigR}\IndexS}}{blue,thick}
\tikzAGCmakeSignal{LSR}{\AGCRightWardSlow{{\SigL}\IndexS}}{blue,thick}
\tikzAGCmakeSignal{LSL}{\AGCLeftWardSlow{{\SigL}\IndexS}}{blue,thick}

\tikzAGCmakeSignal{dsepOne}{\AGCLeftWard{dsep$_1$}}{DarkOrange,thick}
\tikzAGCmakeSignal{dsepTwo}{\AGCRightWard{dsep$_2$}}{DarkOrange,thick}

\JDLvocabulary{\SPEEDSHRINKBACK}{\JDLvocabularyTextXspace{{\AGCspeed}$_{\text{shrink}}^{\text{back}}$\rule{0em}{1em}}}{}{speed of intermediate signals when shrinking from leftward to right ward}
\newcommand{\SPEEDSHRINKBACKvalue}{1/3}

\JDLvocabulary{\SpeedBounceSHRINKBACK}{\JDLvocabularyTextXspace{{\AGCspeed}$_{\text{shrink}}^{\text{bounceBack}}$\rule{0em}{1em}}}{}{speed of intermediate signals when shrinking from leftward to right ward}
\newcommand{\SpeedBounceSHRINKBACKvalue}{3/5}

\tikzAGCmakeSignal{WallSplit}{wall\IndexSplit}{dashed}
\tikzAGCmakeSignal{BounceSBR}{\AGCRightWardSlow{bounce\IndexSB}}{VeryDarkGreen,thick} \tikzAGCmakeSignal{BounceSBL}{\AGCLeftWardSlow{bounce\IndexSB}}{VeryDarkGreen,thick} 
\JDLvocabulary{\SpeedBounceSHRINK}{\JDLvocabularyTextXspace{{\AGCspeed}$_{\text{shrink}}^{\text{bounce}}$\rule{0em}{1em}}}{}{speed of intermediate signals when shrinking from leftward to rightward}
\newcommand{\SpeedBounceSHRINKvalue}{1}

\tikzAGCmakeSignal{BounceSR}{\AGCRightWardSlow{bounce\IndexS}}{VeryDarkGreen,thick} \tikzAGCmakeSignal{BounceSL}{\AGCLeftWardSlow{bounce\IndexS}}{VeryDarkGreen,thick} 
\tikzAGCmakeSignal{BoundSR}{\AGCRightWardSlow{bound\IndexS}}{VeryDarkGrey,thick}
\tikzAGCmakeSignal{BoundSL}{\AGCLeftWardSlow{bound\IndexS}}{VeryDarkGrey,thick}
\tikzAGCmakeSignal{BoundSBR}{\AGCRightWardSlow{bound\IndexSB}}{VeryDarkGrey,thick}
\tikzAGCmakeSignal{BoundSBL}{\AGCLeftWardSlow{bound\IndexSB}}{VeryDarkGrey,thick}

\tikzAGCmakeSignal{OneSBL}{\AGCLeftWardSlow{one\IndexSB}}{DarkBrown,thick}
\tikzAGCmakeSignal{TwoSBL}{\AGCLeftWardSlow{two\IndexSB}}{DarkOrange,thick}
\tikzAGCmakeSignal{LSBL}{\AGCLeftWardSlow{{\SigL}\IndexSB}}{Blue,thick}
\tikzAGCmakeSignal{RSBL}{\AGCLeftWardSlow{{\SigR}\IndexSB}}{Blue,thick}
\tikzAGCmakeSignal{OneSBR}{\AGCRightWardSlow{one\IndexSB}}{DarkBrown,thick}
\tikzAGCmakeSignal{TwoSBR}{\AGCRightWardSlow{two\IndexSB}}{DarkOrange,thick}
\tikzAGCmakeSignal{LSBR}{\AGCRightWardSlow{{\SigL}\IndexSB}}{Blue,thick}
\tikzAGCmakeSignal{RSBR}{\AGCRightWardSlow{{\SigR}\IndexSB}}{Blue,thick}

\tikzAGCmakeSignal{BounceSplitL}{\AGCLeftWard{bounce\IndexSplit}}{VeryDarkGreen, dashed,thick} 
\tikzAGCmakeSignal{BounceSplitOneL}{\AGCLeftWard{bounce$^1\IndexSplit$}}{VeryDarkGreen, dashed,thick} 
\tikzAGCmakeSignal{BoundSPLOne}{bound$^1\IndexSplit$}{VeryDarkGrey, dashed,thick}
\tikzAGCmakeSignal{BoundSOne}{\AGCLeftSide{bound$_1$}}{VeryDarkGrey,thick}\tikzAGCmakeSignal{sepOne}{\AGCRightWard{sep$_1$}}{DarkOrange,thick}
\tikzAGCmakeSignal{sepTwo}{\AGCLeftWard{sep$_2$}}{DarkOrange,thick}
\tikzAGCmakeSignal{sepThree}{\AGCRightWard{sep$_3$}}{DarkOrange,thick}

\tikzAGCmakeSignal{OneSOne}{one$_1$}{DarkBrown,thick}\tikzAGCmakeSignal{TwoSOne}{two$_1$}{DarkOrange,thick}\tikzAGCmakeSignal{LSOne}{l$_1$}{Blue,thick}\tikzAGCmakeSignal{RSOne}{r$_1$}{Blue,thick}
\JDLvocabulary{\SPEEDSPLITUP}{\JDLvocabularyTextXspace{\AGCspeed\!\!$^{\text{up}}_{\text{split}}$}}{}{speed of the upper signal involved in updating parameters after a split}
\newcommand{\SPEEDSPLITUPvalue}{5/3}
\newcommand{\definMSinSide}[4]{\tikzAGCmakeSignal{#1#2}{\AGCLeftRightSide{#3}}{#4}
  \tikzAGCmakeSignal{#1L#2}{\AGCLeftSide{#3}}{#4}
  \tikzAGCmakeSignal{#1R#2}{\AGCRightSide{#3}}{#4}
}

\definMSinSide{SplitUpdtEnd}{}{\SigPreComputeMove}{DarkGreen,thick}\definMSinSide{PreSplitUpdt}{}{updt$^{\text{*}}\IndexSplit$}{DarkGreen,thick}\definMSinSide{SplitUpdtLow}{}{updt$^{\text{lo}}\IndexSplit$}{DarkGreen,thick}\definMSinSide{SplitUpdtHigh}{}{updt$^{\text{hi}}\IndexSplit$}{DarkGreen,thick}\definMSinSide{SplitUpdtBack}{}{updt$^{\text{bck}}\IndexSplit$}{DarkGreen,thick}\definMSinSide{SplitUpdtSet}{}{updt$^{\text{st}}\IndexSplit$}{DarkGreen,thick}\definMSinSide{SplitUpdtSetOne}{}{updt$^{\text{st\_1}}\IndexSplit$}{DarkGreen,thick}\definMSinSide{SplitUpdtLSet}{}{updt$^{\text{lst}}\IndexSplit$}{DarkGreen,thick}\definMSinSide{SplitUpdtRSet}{}{updt$^{\text{rst}}\IndexSplit$}{DarkGreen,thick}
\definMSinSide{SplitUpdtI}{}{updt$^{i}\IndexSplit$}{DarkGreen,thick}\definMSinSide{SplitUpdtISet}{}{updt$^{i\text{st}}\IndexSplit$}{DarkGreen,thick}

\begin{document}

\begin{frontmatter}

  \title{Abstract Geometrical Computation 11: Slanted Firing Squad Synchronisation on Signal Machines}

  \author[1LIFO]{J{\'e}r{\^o}me Durand-Lose\corref{cor1}}
  \ead{jerome.durand-lose@univ-orleans.fr}
  \ead[url]{http\string://www.univ-orleans.fr/lifo/Members/Jerome.Durand-Lose}

  \cortext[cor1]{Corresponding author}
  
  \author[1LIFO]{Aur{\'e}lien Emmanuel}
  \ead{aurelien.emmanuel@univ-orleans.fr}
  
  \address[1LIFO]{Universit{\'e} d'Orl{\'e}ans, INSA Centre Val de Loire, LIFO EA 4022, FR-45067 Orl{\'e}ans, France}

  \begin{abstract}
    \parindent .75cm
    Firing Squad Synchronisation on Cellular Automata is the dynamical synchronisation of finitely many cells without any prior knowledge of their range.
    This can be conceived as a signal with an infinite speed.
    Most of the proposed constructions naturally translate to the continuous setting of signal machines and generate fractal figures with an accumulation on a horizontal line, i.e. synchronously, in the space-time diagram.
    Signal machines are studied in a series of articles named Abstract Geometrical Computation.

    In the present article, we design a signal machine that is able to synchronise/accumulate on any non-infinite slope.
    The slope is encoded in the initial configuration.
    This is done by constructing an infinite tree such that each node computes the way the tree expands.
    
    The interest of Abstract Geometrical computation is to do away with the constraint of discrete space, while tackling new difficulties from continuous space.
    The interest of this paper in particular is to provide basic tools for further study of computable accumulation lines in the signal machine model.
  \end{abstract}


\end{frontmatter}

\section{Introduction}
\label{sec:intro}
The Firing Squad Synchronisation Problem (FSSP) is as follows: a general wants a whole line of riflemen to fire synchronously.
However, communication is only between very close individuals and the number of riflemen is unknown.
There is no global time, but everyone receives ticks simultaneously and repeatedly. 
Moreover, it is supposed that the number of states of each rifleman is finite.
Finite state, discrete space and time, synchrony and uniformity correspond to the Cellular Automata (CA) model.
In this context, the goal is, starting from a single activated cell, to reach a configuration in which all cells of a given region are in a special firing state simultaneously and for the first time.

This problem have been studied from the 1960's \cite{goto62,moore64,waksman66}, with a wide range of construction \cite{culik.ii89,mazoyer87tcs,mazoyer96tcs} and is still active nowadays \cite{luidnel+yunes12acri,luidnel+yunes16acri,luidnel+yunes16jca,luidnel+yunes18,umeo2017,yunes07mcu}.
\phantomsection \label{review:time-minimal-CA}In \cite{mazoyer87tcs}, a six-state minimal time solution is given.
Solutions to this problem also often serve as a tool to solve other problems.
In \cite{culik.ii89} a variation with two generals is used to solve a related complexity problem.
In \cite{NICHITIU2001302}, the election leader problem, which can be seen as a reversed FSSP is solved efficiently for any finite dimension.
In \cite{martin94tcs}, FSSP is used on demand to synchronise computation steps.

The most common solutions involve signals sent at different speeds bouncing onto the edges of the line and sprouting new signals whenever crossing each other in such a way that signals eventually and simultaneously evenly fill the line which triggers the firing.
A tree structure often appears in the construction.
These trees are rooted in the general and the leaves reach the firing riflemen.
Very commonly, the solutions are presented in a continuous setting, then adapted to the discrete space and time of CA; reaching the granularity of space (i.e. cells) indicates when to fire.

It appears natural to consider this problem in the continuous setting of Abstract geometrical computation: the study of computing in an euclidean space with geometrical tools.
It is done using signal machines, which compute by drawing coloured lines which interact in specific ways upon collision.
It lead to a series of articles, for example, isolated accumulations on single point are characterised in \cite{durand-lose11nc-uc}.

\begin{figure}[hbt]
  \scriptsize\newlength{\Height}
  \setlength{\Height}{.22\textwidth}\centerline{\subcaptionbox{space-time diagram\label{fig:random}}{\mbox{}\quad\SetUnitlength{\Height}\begin{tikzpicture}\draw[->] (-1mm,0) -- node[above,sloped] {\scriptsize time} (-1mm,1) ;
        \path (0,0) node[anchor=south west] (A) {\includegraphics[height=\Height]{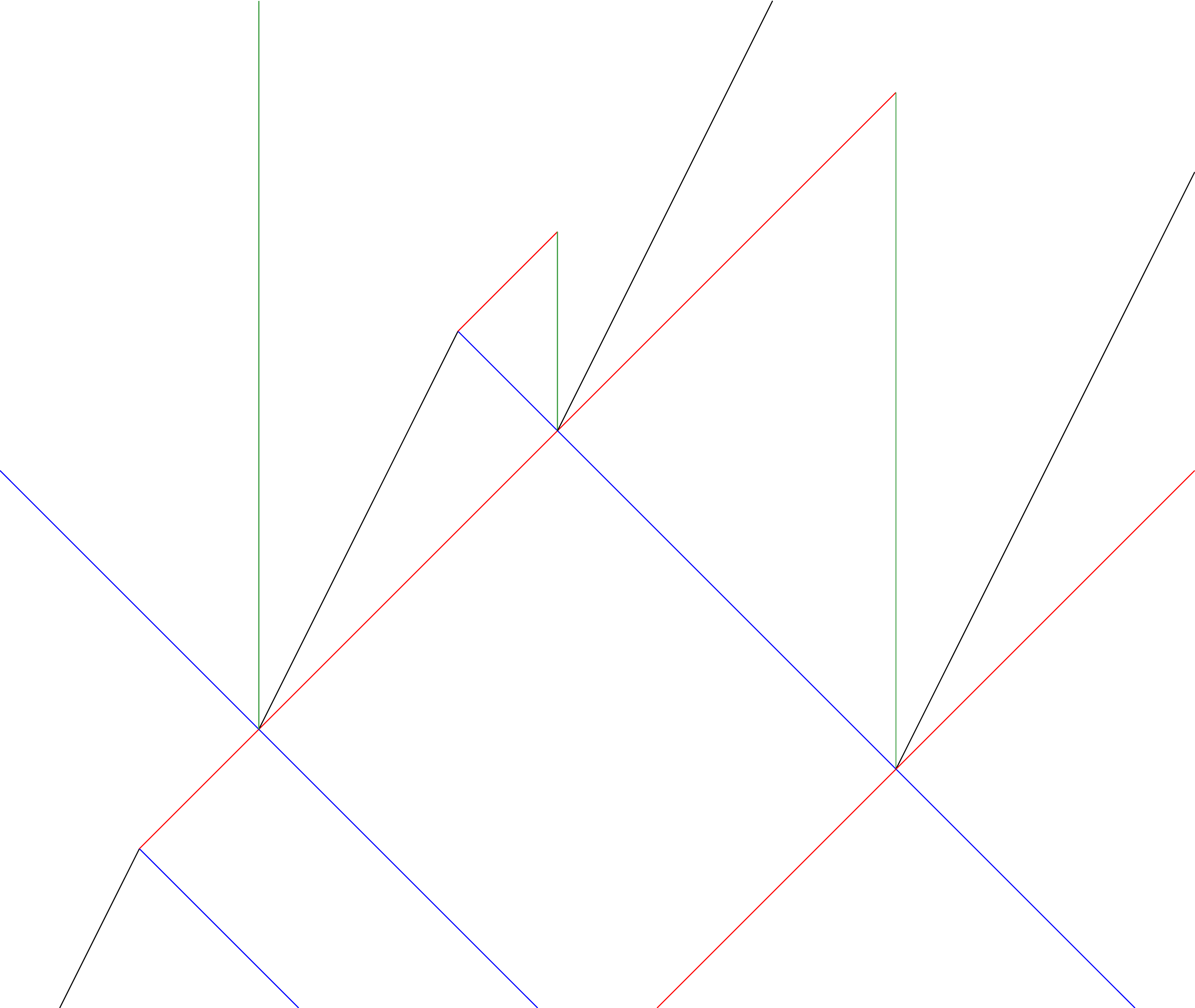}} ;
        \draw[<->] (0,-1mm) -- node[below,sloped] {\scriptsize space} ([yshift=-1mm]A.south east) ;
      \end{tikzpicture}\quad\mbox{}}\addtolength{\Height}{4mm}
    \ \subcaptionbox{isolated accumulation\label{fig:zig-zag}}{\mbox{}\qquad\includegraphics[height=\Height]{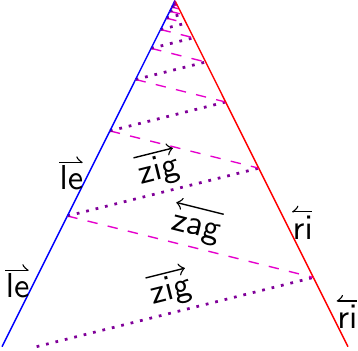}\qquad\mbox{}}\ \subcaptionbox{segment accumulation\label{fig:goto}}{\mbox{}\qquad\qquad\includegraphics[height=\Height]{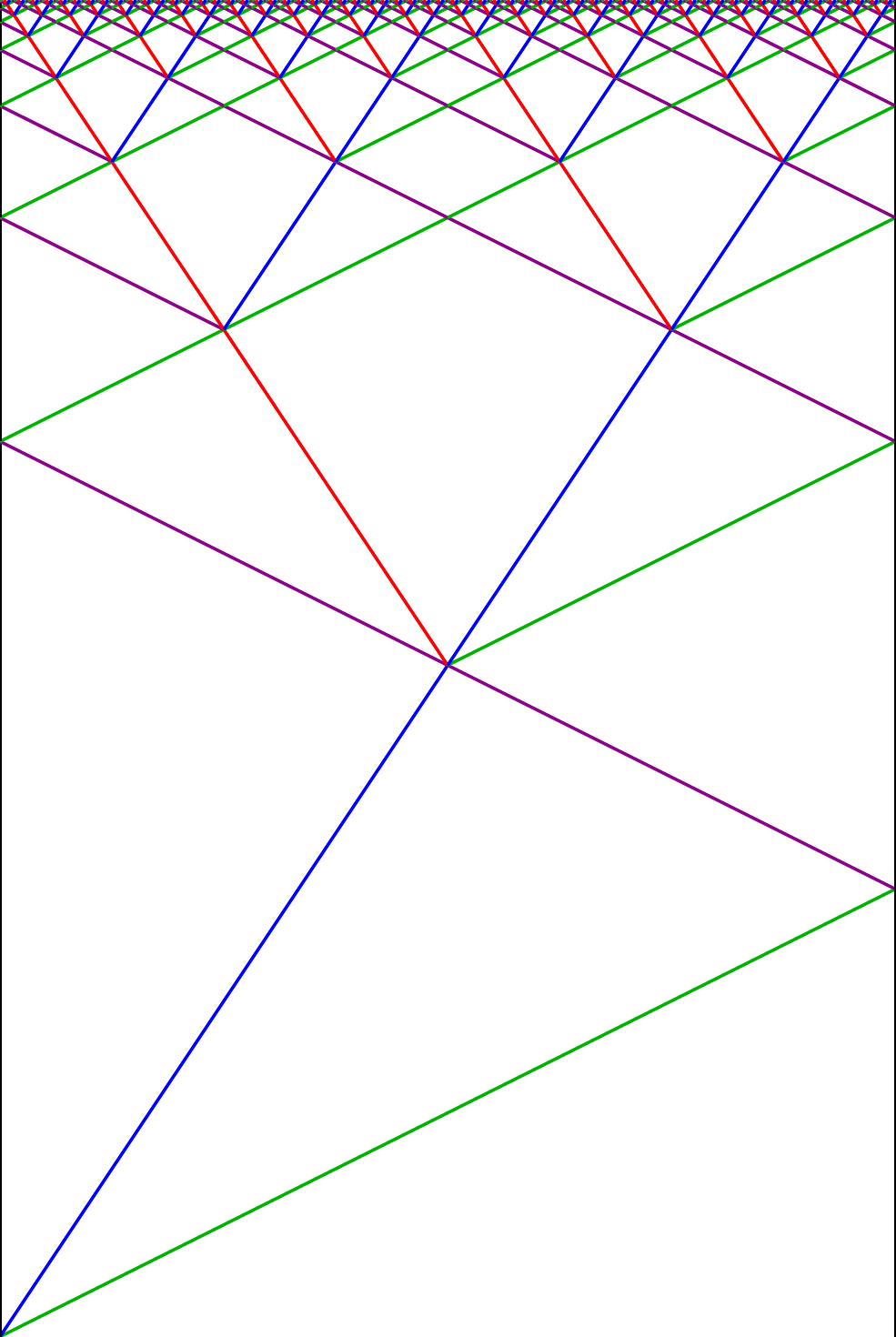}\qquad\qquad\mbox{}}}
  \NTVS
  \caption{Space-time diagrams and basic accumulations.}
  \label{fig:spt+accumulations}
\end{figure}

Signal machines are an abstraction of 1D CA: signals carrying a given label travel at a given speed through a continuous space; any collision between two or more signals results in the vanishing of incoming signals, and the emergence of new ones, according to their labels and to some predefined rules.
\RefFigure{fig:random} provides some space-time diagram of an example signal machine (time elapses upward).
Signal machines are defined by meta-signals that define signals, and collision rules, that define their interactions;
for example, the space-time diagram in \RefFig{fig:zig-zag} was obtained with a machine with meta-signals \SigZig, \SigZag, \SigZZle and \SigZZri,
as well as the collision rules \AGCruleDef{\SigZig,\SigZZri}{\SigZag,\SigZZri} and \AGCruleDef{\SigZag,\SigZZle}{\SigZig,\SigZZle}.

Compared to CA, there is no underlying grid and only signal dynamics is addressed.
The continuous time and space allow phenomena alien to CA: accumulation as illustrated in \RefFig{fig:zig-zag}: there are infinitely many collisions leading to the top of the triangle.
It is already known that forecasting of any accumulation on rational signal machines is highly unpredictable \cite{durand-lose06tamc} and the possible localisation of isolated accumulation of such machines have been characterised \cite{durand-lose11uc}.

Most of the schemes used for FSSP in CA can be defined directly as signals machines (but going from there to a CA is no triviality \cite{besson18phd}).
In CA, the construction stops when the granularity of space is reached.
In the continuous setting of signal machines, there is no such a thing so that the divide and conquer process never stops and generates a fractal.
The result, a horizontal firing line in a space-time diagram, consists of an accumulation line as in \RefFig{fig:goto} (each colour corresponds to one meta-signal, speed magnitudes are 0, 1 or 3, such that red and blue signals form an infinite binary tree).
That is a line of points nearby which an infinity of signal collisions happen (the continuous equivalent of filling a discrete line with signals).
Such a construction with a finite yet unbounded binary tree has been used in \cite{duchier+durand-lose+senot12tamc} to provide unbounded branching to solve PSPACE-complete problems with polynomial depth.

\phantomsection\label{review:slope}The accumulation line in the space-time diagram needs not be horizontal although there are not much research on the matter since, in the more usual discrete settings, synchronisation is the goal and slope is less meaningful.
Achieving a given slope would correspond to riflemen firing in succession at a predefined rate.
Much like the regular FSS problem is about synchrony, our Slanted FSS problem is about coordination.
Its solution could be used to as a tool for problems requiring tasks to be done in a quick succession---too quick for the end of a task to trigger the beginning of the next one;
for example sending self driving cars through an intersection with shorter security distances than possible using uncoordinated human drivers. 

In the present paper, we prove that not only it is possible to generate any segment of any non-infinite slope as an accumulation set but moreover that they can all be generated by the same signal machine. Without loss of generality, we aim at segment with extremities at abscissa $-1$ and $+1$.
The target segment is encoded in the initial configuration: the ordinates (the times) of the extremities are encoded as distance between signals.

The idea is to draw an infinite unary-binary tree in a finite amount of space, halving the size of edges after each node with two children.
The structure of the tree dictates the shape of the accumulation set.
As in \cite{duchier+durand-lose+senot12tamc}, the tree structure conveys some information used to extend the tree: with sign test, addition and subtraction (on real numbers) primitives, each node computes whether to delay or split and updates and broadcasts the information.
These primitives are also part of the linear version of the Blum, Shub and Smale model of computation \cite{blum+shub+smale89bams} witch is related to Signal Machines \cite{durand-lose07cie}.

\phantomsection\label{review:why-ubt}Unary-binary tree is one of the many possible constructions that can fulfil our goal, we chose it for its simplicity, with complete disregard for time minimality or optimisation of the number of meta-signals or collision rules - corresponding notions to states and rule of a cellular automaton.

The algorithm is first implemented on an \emph{augmented signal machine} to allow the signals in the tree to carry an unlimited quantity of information.
This allows to concentrate on the algorithm without dealing with technical details.
Then these augmented signals are encoded as a ray of signals on an ordinary signal machine and needed arithmetic operations are implanted.
This is much more involving because it has to deal with updating the information now encoded with only finitely many possible signals.

The paper is organised as follows.
The relevant definitions are provided in \RefSec{sec:definitions}.
The algorithm is explained in \RefSec{sec:algorithm} and implemented on augmented signal machines in \RefSec{sec:ASMImplementation}.
This is then implemented on ordinary signal machine in \RefSec{sec:implementation}.
\RefSection{sec:conclusion} concludes the paper.

\section{Definitions}
\label{sec:definitions}
\subsection{Signal machines}
A signal machine regroups the definitions of its meta-signals and their dynamics: constant speed outside of collisions and rewriting rules at collisions.

A \emph{signal machine} is a triplet $(\AGCmetaSignalSet,\AGCspeedFun,\AGCruleSet)$ such that: 
\begin{itemizeCompact}
\item \AGCmetaSignalSet is a finite set, whose elements are called \emph{meta-signals};
\item $\AGCspeedFun:\AGCmetaSignalSet \rightarrow \mathbb{R}$ is the \emph{speed function}  (each meta-signal has a constant speed);
\item \AGCruleSet is a finite set of \emph{collision rules}; a collision rule \AGCrule can be written $\AGCruleIn \rightarrow \AGCruleOut$, and consists in an \emph{input set} \AGCruleIn and an \emph{output set} \AGCruleOut of meta-signals of distinct speeds, with \AGCruleIn containing at least two meta-signals.
  \AGCruleSet is deterministic: $\AGCrule \neq \AGCrulePrime$ implies that $\AGCruleIn \neq \AGCrulePrimeIn$.
\end{itemizeCompact}

In the example in \RefFig{fig:zig-zag}, there are four meta-signals:
\SigZag,
\SigZZri, 
\SigZZle, and
\SigZig
of speeds $-1$, $-1/2$,  $1/2$, and $1$ respectively.
The collision rules are
\AGCruleDef{\SigZig,\SigZZri}{\SigZag,\SigZZri}
and
\AGCruleDef{\SigZZle,\SigZag}{\SigZZle,\SigZig}.

A \emph{configuration}, \AGCconfiguration, is a mapping from the points of the real line to either a meta-signal, a rule or the value \AGCextendedValueVoid (indicating that there is nothing there).
There are finitely many non-\AGCextendedValueVoid locations in any \emph{initial configuration}.
There are 3 signals in the initial configuration in \RefFig{fig:zig-zag}, from left to right: 
\SigZZle, \SigZig, and \SigZZri.

A \emph{space-time diagram} is the collection of configurations as time elapses.
It forms a two dimensional picture (time is elapsing upwards in the figures).
It is a function from $\RealSet\times\RealSet^{+}\mapsto \AGCmetaSignalSet\cup\AGCruleSet\cup\{\AGCextendedValueVoid\}$.

\phantomsection \label{review:signal}A \emph{signal} of a space-time diagram is a maximal segment or half-line mapped by the space-time diagram to a meta-signal \AGCmetaSignal and of inverse slope $\AGCspeedFun(\AGCmetaSignal)$ (signals cannot be horizontal).
Meta-signals can be thought of as type, and signals as instances of them.
The meta-signal of a signal is also called its \emph{type} for the sake of concision.

A space-time diagram follows the rules: 
\begin{itemizeCompact}
\item any point of the space-time diagram that maps to a meta-signal \AGCmetaSignal belongs to a signal which verifies:
  \begin{itemizeCompact}
  \item its starting location is either a rule \AGCrule with $\AGCmetaSignal \in \AGCruleOut$ or a point of the initial configuration mapping to \AGCmetaSignal and
  \item its end, if any, is a rule \AGCrule with $\AGCmetaSignal \in \AGCruleIn$;
  \end{itemizeCompact}
\item any point of the space-time diagram with a collision of rule \AGCrule is at:
  \begin{itemizeCompact}
  \item the starting end of one signal of type \AGCmetaSignal for every \AGCmetaSignal in \AGCruleOut and
  \item the ending end of one signal of type \AGCmetaSignal for every \AGCmetaSignal in \AGCruleIn (if not in the starting configuration);
  \end{itemizeCompact}

  \phantomsection\label{review:path}\item there is no infinite time-backward continuous path/sequence of signals each of whose start is the end of the next.
\end{itemizeCompact}

This ensures that signals propagate with uniform speed, collision rules are properly applied and no signal nor collision appears out of the blue.
In particular we do not want anything appearing from accumulation points as defined below.

\subsection{Accumulation points}

In the example in \RefFig{fig:zig-zag}, there is a special point on top such that there are infinitely many signals and collisions leading to it.
The signal machine does not provide any definition for what happens there so that the space-time diagram cannot be defined for all times and locations.
We extend the definition as follows: a space-time diagram is a partial function from $\RealSet\times\RealSet^{+}$ to  $\AGCmetaSignalSet\cup\AGCruleSet\cup\{\AGCextendedValueVoid,\AGCextendedValueAcc\}$ where \AGCextendedValueAcc is a new symbol introduced to designate accumulation points (and only such points).
Outside of accumulation points, the constrains are as before but signals may also end in accumulations.
Any accumulation point must be the forward limit of an infinite sequence of collision points.

The example in \RefFig{fig:zig-zag} is the simplest example of accumulation.
There is only one point valued \AGCextendedValueAcc; the rest of the space-time diagram is  \AGCextendedValueVoid.
In the example of \RefFig{fig:goto}, a connected accumulation set (a line segment) is generated.

\subsection{Augmented signal machines}

\emph{Augmented Signal Machines (ASM)} are natural extensions of signal machine where signals are allowed to carry analog information.

An \emph{augmented signal machine} is a triplet $(\AGCmetaSignalSet,\AGCspeedFun,\AGCruleSet)$ such that: 
\begin{itemizeCompact}
\item $\AGCmetaSignalSet$ is a finite set of \emph{meta-signals};
\item $\AGCspeedFun:\AGCmetaSignalSet \rightarrow \mathbb{R}$ is the \emph{speed function};
\item each $\AGCmetaSignal$ in \AGCmetaSignalSet has an associated set, called \emph{domain} and noted $\AGCaugSignalDomainFun_{\AGCmetaSignal}$: signals may carry information depending on their type;
\item the set of \emph{augmented meta-signals} is $\AGCaugmentedMetaSignalSet = \{(\AGCmetaSignal, x) | \AGCmetaSignal \in \AGCmetaSignalSet, x \in \AGCaugSignalDomainFun_{\AGCmetaSignal}\}$.
  The speed of an augmented meta-signal is simply the speed of its corresponding meta-signal;
\item \AGCruleSet is a set of \emph{collision rules}; a collision rule \AGCrule can be written $\AGCruleIn \rightarrow \AGCruleOut$, and consists in an \emph{input set} \AGCruleIn and an \emph{output set} \AGCruleOut of augmented meta-signals of distinct speeds, with \AGCruleIn containing at least two augmented meta-signals.
  \AGCruleSet is deterministic.
\end{itemizeCompact}

Configurations, diagrams, augmented signals and accumulation points are defined the same way as for regular signal machines, replacing meta-signals by augmented meta-signals.

A signal machine is a special case of augmented signal machine where every $\AGCaugSignalDomainFun_{\AGCmetaSignal}$ is a singleton (or a finite set).
In this paper, domains are either singletons or subsets of $\RealSet^2$.
We thus note augmented meta-signals $\AGCmetaSignal$ when $\AGCaugSignalDomainFun_{\AGCmetaSignal}$ is a singleton, and otherwise we note $\AGCaugmentedMetaSignal$ instead of $(\AGCmetaSignal, (\InputS, \InputT))$.

Since there are potentially infinitely many rules for an augmented signal machines, we define several of them at a time through parameterized patterns:
$X \xrightarrow{C} Y$
where $X$ is a left member of a rule containing free variables, $C$ is a condition on them and $Y$ is a right member which might also depend on them.
Concrete examples of rule pattern are given in \RefSec{sec:ASMImplementation}.

\subsection{Slanted Firing Squad Synchronization Problem}
\label{subsection:SFSS}

The goal of the \emph{Firing Squad Synchronization Problem} (FSSP) on signal machines is to design a machine and an initial configuration which create an horizontal line as the set of accumulation points of the corresponding space-time diagram.

Formally, a solution of the FSSP is a signal machine together with an initial configuration composed of two border signals, a sequence of signals\,---called the general---\,along with the prescribed distance between them near the left border.
If the width of the general is small enough relative to distance between border signals, there exists a time $\TimeCoordinate_l$ such that, in the resulting space-time diagram, the set of accumulation point is the horizontal segment of equation $\TimeCoordinate = \TimeCoordinate_l$\,---within the boundaries.

The goal of the \emph{Slanted Firing Squad Synchronization Problem} (SFSSP) for signal machines is to design a machine and an initial configuration which creates a slanted line as the set of accumulation points of the corresponding space-time diagram.
In other words, looking at the firing line over time, we want the firing to happen in succession, as an apparent dot moving at a constant, specified speed.
Since space-time diagram is a central piece of signal machines, we prefer the image of and the vocabulary related to a slanted line in a space-time diagram over those of a virtual moving dot.

Formally, a \emph{solution} of the Slanted FSSP of slope \Slope is again a signal machine and an initial configuration, such that
\begin{itemizeCompact}
\item the initial configuration is composed of 2 static bounding signals (to delimit the firing line), one on the left border of the general, one far beyond the right border;
\item for all big enough ratio of width between the line and the general, there exists a time $\TimeCoordinate_l$ such that, in the resulting space-time diagram, the set of accumulation points is the slanted line segment of equation $\TimeCoordinate = \TimeCoordinate_l + \Slope \times \SpaceCoordinate$ within the boundaries.
\end{itemizeCompact}

Conversely, instead of aiming at a big enough line, we can shrink the general.
This is our approach in the rest of the paper and we fix the boundaries at $-1$ and $1$.

Additionally and informally, a \emph{universal solution} to the SFSSP is a single signal machine and a way to encode any given slope \Slope into a general so as to solve the corresponding SFSSP.

In this paper, instead of aiming at a slope \Slope, we aim at a segment $[(-1,\InputS), (1,\InputT)]$ with $\InputS, \InputT \geq 1$.
This allows to solve for any slope, and gives perspective about drawing arbitrary segment or curve in the space-time diagram, while foregoing time-optimality.

\section{Algorithm and ASM implementation}
\label{sec:algorithm}
\subsection{General scheme and algorithm}

\begin{figure}[htb]
  \centering\scriptsize\SetUnitlength{.0085\textwidth}\newcommand{\BorderR}{11}
  \newcommand{\BorderL}{-\BorderR}
  \newcommand{\BorderB}{0}
  \newcommand{\BorderU}{\BorderR}
  \subcaptionbox{parameters\label{fig:parameters}}{\centering\tiny\begin{tikzpicture}
      \path[use as bounding box] (\BorderL-1-1,-.75) rectangle (\BorderR+1+1, \BorderU*2.25) ;
      \draw[thick](\BorderL,\BorderB) -- (\BorderL,1.25*\BorderU);
      \draw[thick](\BorderR,\BorderB) -- (\BorderR,2.25*\BorderU);
      \fill (0,0) circle (.2) node [above] {(\InputS,\InputT)};
      \begin{scope}[very thin,draw=VeryDarkGrey,inner sep=.4mm]
  	\draw[<->] (0,-1) -- node[above] {1} +(\BorderR,0) ;
  	\draw[<->] (0,-1) -- node[above] {1} +(\BorderL,0) ;
  	\draw[->] (\BorderL-1,0) -- node[left] {\InputS} +(0,1.25*\BorderU) ;
  	\draw[->] (\BorderR+1,0) -- node[right] {\InputT} +(0,2.25*\BorderU) ;
      \end{scope}
      \draw[densely dotted] (\BorderL,1.25*\BorderU) -- node [sloped,above]{\AGCextendedValueAcc target} (\BorderR,2.25*\BorderU) ;
    \end{tikzpicture}}
  \quad\subcaptionbox{tree\label{fig:tree}}{\newcommand{\TTrree}[3]{\draw[Red,thick] (#1) -- +(-#2,#3) ;
      \draw[Blue,thick] (#1) -- +(#2,#3) ; 
      \draw[thick,densely dotted] (#1) ++(#2,#3) -- +(0,#3) ;
    }
    \centering\pgfmathsetmacro\MiddleH{(\BorderL + \BorderR) / 2}
    \pgfmathsetmacro\MiddleV{(\BorderU + \BorderB) / 2}
    \SetUnitlength{.5\unitlength}\begin{tikzpicture}
      \TTrree{0,0}{\BorderR}{\BorderU}
      \TTrree{-\BorderR,\BorderU}{\BorderR/2}{\BorderU/2}
      \TTrree{\BorderR,2*\BorderU}{\BorderR/2}{\BorderU/2}
      \TTrree{-1.5*\BorderR,1.5*\BorderU}{\BorderR/4}{\BorderU/4}
      \TTrree{-.5*\BorderR,2*\BorderU}{\BorderR/4}{\BorderU/4}
      \TTrree{.5*\BorderR,2.5*\BorderU}{\BorderR/4}{\BorderU/4}
      \TTrree{1.5*\BorderR,3*\BorderU}{\BorderR/4}{\BorderU/4}
    \end{tikzpicture}}\quad\subcaptionbox{\StepDelay\label{fig:delay_simple}}{\centering\pgfmathsetmacro\MiddleH{(\BorderL + \BorderR) / 2}
    \pgfmathsetmacro\MiddleV{(2*\BorderU + \BorderB) / 3}
    \begin{tikzpicture}
      \path[use as bounding box] (\BorderL-1,-.75) rectangle (\BorderR+1, \BorderU) ;
      \begin{scope}[very thin,draw=VeryDarkGrey,inner sep=.4mm]
        \draw[<->] (0,-1) -- node[above] {\tiny 1} +(\BorderR,0) ;
        \draw[<->] (-\BorderR,-1) -- node[above] {\tiny 1} +(\BorderR,0) ;
        \draw[<->] (\BorderR+1,0) -- node[right] {\tiny 1} +(0,\BorderU) ;
      \end{scope}
      \draw[thick] (\BorderL,\BorderB) -- (\BorderL, \BorderU);
      \draw[thick] (\BorderR,\BorderB) -- (\BorderR, \BorderU);\draw[VeryDarkGreen,thick](\MiddleH,\BorderB) -- (\BorderR, \MiddleV);\draw[Purple,thick](\BorderR, \MiddleV) -- (\MiddleH,\BorderU);\draw[thick,densely dotted](\MiddleH,\BorderB) -- (\MiddleH,\BorderU);\end{tikzpicture}}\qquad\subcaptionbox{\StepSplit\label{fig:split_simple}}{\centering\renewcommand{\BorderU}{5.5}
    \pgfmathsetmacro\Width{\BorderR - \BorderL}
    \pgfmathsetmacro\Height{\BorderU - \BorderB}
    \begin{tikzpicture}
      \path[use as bounding box] (\BorderL-1,-.75) rectangle (\BorderR+1, \BorderU) ;
      \begin{scope}[very thin,draw=VeryDarkGrey,inner sep=.4mm]
        \draw[<->] (0,-1) -- node[above] {\tiny 1} +(\BorderR,0) ;
        \draw[<->] (-\BorderR,-1) -- node[above] {\tiny 1} +(\BorderR,0) ;
        \draw[<->] (-\BorderR-1,0) -- node[left] {\tiny \!1\!/\!2} +(0,\BorderU) ;
      \end{scope}
      \draw[thick] (\BorderL,\BorderB) -- (\BorderL, \BorderU);
      \draw[thick] (\BorderR,\BorderB) -- (\BorderR, \BorderU);\path[thick] (\BorderL + \Width / 2,\BorderB) \CoorNode {init} ;
      \path[thick] (\BorderL + \Width / 4,\BorderU) \CoorNode {left} ;
      \path[thick] (\BorderR - \Width / 4,\BorderU) \CoorNode {right} ;
      \draw[Purple,thick] (init) -- (\BorderL, \BorderB + 2 * \Height / 3);\draw[DarkGreen,thick] (\BorderL, \BorderB + 2 * \Height / 3) -- (left);\draw[red,thick] (init) -- (left);\draw[Purple,thick] (\BorderR, \BorderB + 2 * \Height / 3) -- (right);\draw[DarkGreen,thick] (init) -- (\BorderR, \BorderB + 2 * \Height / 3) ;
      \draw[blue,thick] (init) -- (right);\draw[thick] (init) -- (\BorderL + \Width / 2, \BorderU); \end{tikzpicture}}\NTVS
  \caption{Parameters, unary-binary tree, and elementary steps.}
\end{figure}
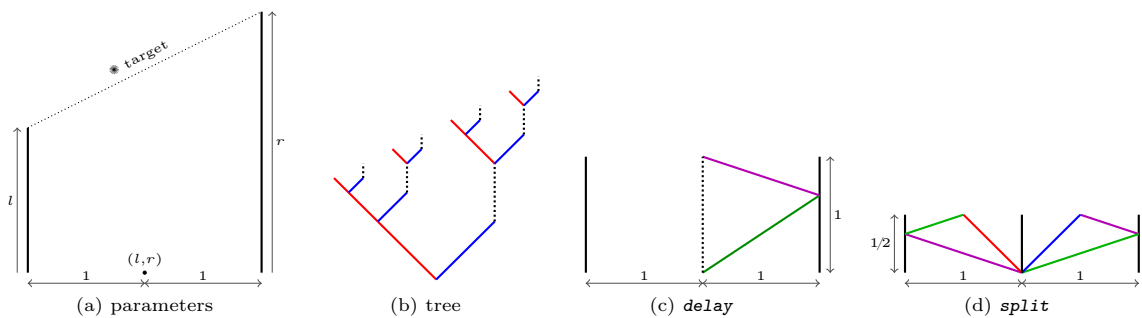

The target segment and the parameters are depicted in \RefFig{fig:parameters} : parameters \InputS and \InputT are the ordinates of the extremities of the target segment.
To be valid, the parameters have to be both at least $1$.

The algorithm relies on a very simple divide and conquer strategy: to \emph{draw}\,---accumulate on---\,the segment $[(-1,\InputS), (1,\InputT)]$, it suffices be able, depending on the situation, to either:
\begin{itemizeCompact}
\item draw the segment $[(-1,\InputS-1), (1,\InputT-1)]$ translated one unite of distance up, or
\item draw the segments $[(-1, \InputS - 1/2),(0, (\InputS + \InputT)/2 - 1/2)]$ and $[(0, (\InputS + \InputT)/2 - 1/2), (1, \InputT - 1/2)]$ translated to have their roots at $(-1/2,1/2)$ and $(1/2,1/2)$ respectively.
  It is scaled by $1/2$ yielding the formula in \RefAlgo{code:aug-rule}.
\end{itemizeCompact}

\RefFigure{fig:tree} illustrates this infinite recursion, with vertical segments drawn as dotted and black, and diagonal segments of slope $+-1$ being drawn as red or blue, depending whether they go to the left or right (ascending) respectively.
This colour code applies to section \ref{sec:ASMImplementation} as well.
In the one entwined in \RefFig{fig:goto} (looking only at red and blue signals), each node always have two descendants.
The unary steps are used to provide delays and thus slope.
Each (sub-)tree is bounded by two motionless signals as in figures \ref{fig:delay_simple} and \ref{fig:split_simple}.
When an update is carried out, the collision happens exactly at the middle of the bounds which are considered 2 units of space apart.

The following primitives are used to direct the dynamics in the algorithm:
\begin{Description}
\item[\StepDelay($\InputS, \InputT$)]: the current node is of degree 1.
  Its single child is 1 unit up and the algorithm is prompted with parameters $(\InputS, \InputT)$. In \RefFig{fig:tree} a \StepDelay corresponds to a dotted connections 
  and
\item[\StepSplit(($\InputS, \InputT$),($\InputS', \InputT'$))]: the current node is of degree 2.
  The left child is at $(-1/2,1/2)$ relatively to the current node, the right child at $(1/2,1/2)$.
  In \RefFig{fig:tree}, the links between the node and its children are red and blue.
  At each child, the algorithm is prompted with the scale reduce by a factor 2 and with the parameters $(\InputS, \InputT)$ for the left one, and $(\InputS', \InputT')$ for the right one.
\end{Description}

\RefFigure{fig:delay_simple} outlines how a delay can be implemented by an augmented signal machine, with two auxiliary signals bouncing off the boundary at the correct speeds ($3/2$ for green and $-3$ for purple) to wait the correct amount of time.
\RefFigure{fig:split_simple} outlines how a split can be implemented, with auxiliary green signals of speed $3$, and purple $-3$.
Altogether, it yields the algorithm \ref{code:aug-rule} where the tree is initialised with the ``top level'' \InputS and \InputT for the targeted final accumulation segment.
The strategy is to delay whenever possible, i.e. when both extremities are at least 2 units above.

\begin{listing}[hbt]
  \centering{\begin{minipage}{.8\linewidth}\small
\begin{minted}{python}
if 2 <= l and 2 <= r :                  # time += 1
        delay ( l - 1 , r - 1 )                            #            remove 1
else :                                         # time += 1/2
        split ( ( l + l - 1 , r + l - 1 ) ,               # add l then remove 1
                ( l + r - 1 , r + r - 1 ) )               # add r then remove 1
\end{minted}
    \end{minipage}}
  \NTVS
  \caption{Code for augmented collision rules.}
  \label{code:aug-rule}
\end{listing}

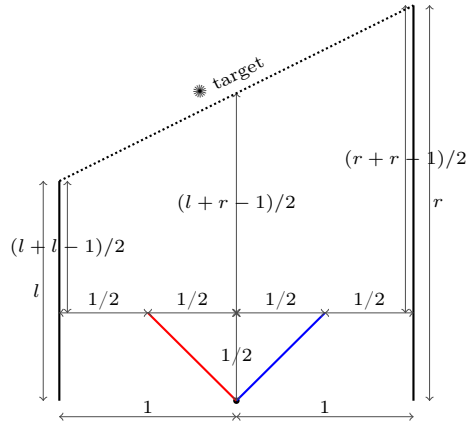
\begin{figure}[htb]
  \centering\scriptsize\SetUnitlength{2ex}\newcommand{\BorderR}{11}
  \newcommand{\BorderL}{-\BorderR}
  \newcommand{\BorderB}{0}
  \newcommand{\BorderU}{\BorderR}
  \newcommand{\HalfUnit}{\BorderR/2}
  
  \begin{tikzpicture}
    \path[use as bounding box] (\BorderL-1-1.5,-.5) rectangle (\BorderR+1+1.5, \BorderU*2.25) ;
    \draw[thick] (\BorderL,\BorderB) -- (\BorderL,1.25*\BorderU);
    \draw[thick] (\BorderR,\BorderB) -- (\BorderR,2.25*\BorderU);
    \fill (0,0) circle (.2);
    \draw[Blue,thick](0,\BorderB) -- (\HalfUnit, \HalfUnit);
    \draw[red,thick](0,\BorderB) -- (-\HalfUnit, \HalfUnit);
    \begin{scope}[very thin,draw=VeryDarkGrey, inner sep=.4mm]
      \draw[<->] (0,-1) -- node[above] {1} +(\BorderR,0) ;
      \draw[<->] (0,-1) -- node[above] {1} +(\BorderL,0) ;
      \draw[<->] (\BorderL-1,0) -- node[left] {\InputS} +(0,1.25*\BorderU) ;
      \draw[<->] (\BorderR+1,0) -- node[right] {\InputT} +(0,2.25*\BorderU) ;
      \draw[<->] (0, \HalfUnit) -- node[above] {1/2} +(\HalfUnit,0) ;
      \draw[<->] (\HalfUnit, \HalfUnit) -- node[above] {1/2} +(\HalfUnit,0) ;
      \draw[<->] (-\HalfUnit, \HalfUnit) -- node[above] {1/2} +(\HalfUnit,0) ;
      \draw[<->] (\BorderL, \HalfUnit) -- node[above] {1/2} +(\HalfUnit,0) ;
      \draw[<->] (0, 0) -- node {1/2} +(0, \HalfUnit);
      \draw[<->] (\BorderL+1/2, \HalfUnit) -- node {$(\InputS + \InputS - 1) / 2$} +(0, 1.25* \BorderU - \HalfUnit);
      \draw[<->] (0, \HalfUnit) -- node {$(\InputS + \InputT - 1) / 2$} +(0, 1.75* \BorderU - \HalfUnit);
      \draw[<->] (\BorderR-1/2, \HalfUnit) -- node {$(\InputT + \InputT - 1) / 2$} +(0, 2.25* \BorderU - \HalfUnit);
    \end{scope}
    \draw[densely dotted,thick] (\BorderL,1.25*\BorderU) -- node [sloped,above]{\AGCextendedValueAcc target} (\BorderR,2.25*\BorderU) ;
  \end{tikzpicture}
  \NTVS
  \caption{\Split update formula illustrated.}
  \label{fig:update-formula}
\end{figure}

\RefFigure{fig:update-formula} illustrates the update formula after a split:
the new target heights are shorter by $1/2$, and are multiplied by $2$ to account for the new scale.

Nodes are called \StepSplit or \StepDelay depending on the used primitive.
There is no end to the recursion: the constructed unary-binary tree is infinite to produce the whole accumulations segment.

\subsection{Correctness}

The \StepSplit nodes play a key role in the demonstration of the correctness of the algorithm.
The \emph{depth} of a collision is the number of its ancestors of degree $2$ (i.e. \StepSplit nodes).
From now on, let $\CollisionDepth$ denotes the depth of a given collision and $\Slope$ denotes the slope of the targeted segment, that is $(\InputTInitial - \InputSInitial) / 2$ where $(\InputTInitial,\InputSInitial)$ denotes the parameter at the root of the tree.

\begin{lemma}[Invariants]\label{lem:Invariants}
  \RefAlgorithm{code:aug-rule} satisfies the following invariants:
  \begin{Enumerate}
  \item $1\leq\InputS$ and $1\leq \InputT$ (the parameters remain valid);
  \item $(\InputT - \InputS) / 2=\Slope$ (slope is preserved);
  \item the boundary points of the current targeted segment are on the initially targeted segment:
    assume the initial parameters are $\InputSInitial$ and $\InputTInitial$, and consider a collision happening at $(\SpaceCoordinate, \TimeCoordinate)$, with parameters \InputS, \InputT.
    The points $(\SpaceCoordinate, \TimeCoordinate) + 2^{-\CollisionDepth} (-1,\InputS)$ and $(\SpaceCoordinate, \TimeCoordinate) + 2^{-\CollisionDepth} (+1,\InputT)$ are on the initial targeted segment.
    That is the segment with extremities $(-1,\InputSInitial)$ and $(1,\InputTInitial)$
  \end{Enumerate}
\end{lemma}

\begin{proof}
  The first two invariants are straightforward to prove from \RefAlgo{code:aug-rule}.
  Let us consider a node at $(\SpaceCoordinate, \TimeCoordinate)$ with parameters $\InputS, \InputT$ (both at least $1$).
  Let us prove the last invariant by induction.
  It is true for the root node.
  
  After a \StepDelay node: $(\SpaceCoordinate - 2^{-\CollisionDepth}, (\TimeCoordinate + 2^{-\CollisionDepth}) + 2^{-\CollisionDepth} (\InputS - 1)) = (\SpaceCoordinate - 2^{-\CollisionDepth}, \TimeCoordinate + 2^{-\CollisionDepth} \InputS)$ and similarly for \InputT: the targeted segment extremities are unchanged.
  
  After a \StepSplit node: $(\SpaceCoordinate - 2^{-(\CollisionDepth + 1)} - 2^{-(\CollisionDepth + 1)}, (\TimeCoordinate + 2^{-(\CollisionDepth + 1)}) + 2^{-(\CollisionDepth + 1)} .2.(\InputS-1/2)) = (\SpaceCoordinate - 2^{-\CollisionDepth}, \TimeCoordinate + 2^{-\CollisionDepth}\InputS)$: the left extremity of the left new branch matches the left extremity of the formerly targeted segment. 
  Since $(\SpaceCoordinate - 2^{-(\CollisionDepth + 1)} + 2^{-(\CollisionDepth + 1)}, (\TimeCoordinate + 2^{-(\CollisionDepth + 1)}) + 2^{-(\CollisionDepth + 1)} .2.((\InputS+\InputT)/2-1/2) = (\SpaceCoordinate, \TimeCoordinate + 2^{-\CollisionDepth} (\InputS + \InputT)/2)$, the right extremity of the left new branch matches the middle of the formerly targeted segment.
  Proof is the same for the right branch.
\end{proof}

\begin{lemma}\label{lem:Split_Col}
  A \StepSplit collision at $(\SpaceCoordinate, \TimeCoordinate)$ happens before the point on the targeted segment at the same spatial location, but not sooner than $2^{-\CollisionDepth} (2 + |\Slope|)$ before.
  In other words:
  \NTVS
  \begin{equation*}
    \TimeCoordinate + 2^{-\CollisionDepth} (\InputS + \InputT) / 2 - 2^{-\CollisionDepth} (2 + |\Slope|)
    \ \leq\
    \TimeCoordinate
    \ \leq\
    \TimeCoordinate + 2^{-\CollisionDepth} (\InputS + \InputT) / 2
    \enspace.
  \end{equation*}
\end{lemma}

\begin{proof} 
  The second inequality is obvious.
  For the first one, we can observe that:
  \begin{math}
    (\InputS + \InputT)/2 = \min \{\InputS, \InputT\} + |\Slope|
  \end{math}.  
  Indeed, we have:
  \begin{Eqnarray}
    2 \min \{\InputS, \InputT\} + |\InputT - \InputS|
    & = & 2 \min \{\InputS, \InputT\} + \max \{\InputS, \InputT\} - \min \{\InputS, \InputT\}
    \\
    & = & \min \{\InputS, \InputT\} + \max \{\InputS, \InputT\} = \InputS + \InputT
    \enspace.
  \end{Eqnarray}

  We can then write:
  \begin{Eqnarray}
    \TimeCoordinate
    & = & \TimeCoordinate + 2^{-\CollisionDepth} (\InputS + \InputT) / 2 - 2^{-\CollisionDepth} (\min \{\InputS, \InputT\} + |\Slope|)
    \enspace.
  \end{Eqnarray}

  Finally, since it is a \StepSplit node, $\min \{\InputS, \InputT\} \leq 2$, yielding the desired inequality.
\end{proof}

Let us note that any infinite branch of the tree contains infinitely many \StepSplit.
There are always infinitely many branching on both sides.

\begin{theorem}[Correctness of the algorithm]
  Given inputs \InputSInitial and \InputTInitial no lesser than $1$, the algorithm draws a tree whose closure is the segment $[(-1,\InputSInitial), (1,\InputTInitial)]$.
\end{theorem}

\begin{proof}
  Let $\SpaceCoordinate$ be in $[-1,1]$. Our goal is to prove that there is an accumulation at $(\SpaceCoordinate, (\InputSInitial + \InputTInitial)/2 + \Slope \SpaceCoordinate)$.
  
  We remark that the set of abscissae of \StepSplit nodes is exactly the set of dyadic rationals comprised strictly between $-1$ and $1$, which is dense in $[-1,1]$. Indeed, these are the number written $\Sigma_1^d a_i 2^{-i}$ with the $a_i$s in $\{+1,-1\}$. 
  
  Thus, there is a sequence of nodes $(\SpaceCoordinate_{\CollisionDepth}, \TimeCoordinate_{\CollisionDepth})_{\CollisionDepth \in \NaturalSet}$ such that:
  \begin{Enumerate}
  \item $(\SpaceCoordinate_{\CollisionDepth}, \TimeCoordinate_{\CollisionDepth})$ is a \StepSplit collision point of depth $\CollisionDepth$.
  \item $(\SpaceCoordinate_{\CollisionDepth})$ converges, with limit $\SpaceCoordinate$. We can even take $|\SpaceCoordinate_{\CollisionDepth} - \SpaceCoordinate| < 2^{-\CollisionDepth}$, and we do, for convenience.
  \end{Enumerate}

  \begin{Eqnarray}
    |\TimeCoordinate_{\CollisionDepth} - ((\InputSInitial + \InputTInitial)/2 + \Slope \SpaceCoordinate)|
    &\leq& |\TimeCoordinate_{\CollisionDepth} - ((\InputSInitial + \InputTInitial)/2 + \Slope \SpaceCoordinate_{\CollisionDepth})|
    \\
    &&+ \  |((\InputSInitial + \InputTInitial)/2 + \Slope \SpaceCoordinate_{\CollisionDepth}) - ((\InputSInitial + \InputTInitial)/2 + \Slope \SpaceCoordinate)|
    \\
    & \leq & |\TimeCoordinate_{\CollisionDepth} - ((\InputSInitial + \InputTInitial)/2 + \Slope \SpaceCoordinate_{\CollisionDepth})| + | \Slope (\SpaceCoordinate_{\CollisionDepth} - \SpaceCoordinate)| 
    \\
    &\leq& |\TimeCoordinate_{\CollisionDepth} - ((\InputSInitial + \InputTInitial)/2 + \Slope \SpaceCoordinate_{\CollisionDepth})| + | \Slope | \times 2^{-\CollisionDepth}
  \end{Eqnarray}

  then,  because of lemmas \ref{lem:Split_Col} and \ref{lem:Invariants},

  \begin{Align}
    -2^{-\CollisionDepth} (2 + |\Slope|) \leq \TimeCoordinate_{\CollisionDepth} - ((\InputSInitial + \InputTInitial)/2 + \Slope \SpaceCoordinate_{\CollisionDepth}) \leq 0 
  \end{Align}
  which implies
  \begin{Align} 
    |\TimeCoordinate_{\CollisionDepth} - ((\InputSInitial + \InputTInitial)/2 + \Slope \SpaceCoordinate_{\CollisionDepth})| \leq 2^{-\CollisionDepth} (2 + |\Slope|)
    \enspace .
  \end{Align}
  
  Putting things together, we have:
  \begin{Align} 
    |\TimeCoordinate_{\CollisionDepth} - (\InputSInitial + \InputTInitial)/2 + \Slope \SpaceCoordinate| \leq 2^{-\CollisionDepth} (2 + |\Slope| + | \Slope |)
    x\xrightarrow[d\rightarrow + \infty]{} 0
    \enspace .
  \end{Align}

  Proving that $(\TimeCoordinate_{\CollisionDepth})$ converges, with limit $(\InputSInitial + \InputTInitial)/2 + \Slope \SpaceCoordinate$. That proves $(\SpaceCoordinate, (\InputSInitial + \InputTInitial)/2 + \Slope \SpaceCoordinate)$ is an accumulation point.
  
  Points above the targeted segment are not accumulation points as no node is above it, as shows \RefLem{lem:Split_Col} (\StepDelay nodes have to have a \StepSplit heir above them, eventually).
  For a point below the accumulated segment, say by a amount $\Duration$, then, according to \RefLem{lem:Split_Col}, there's a depth $\CollisionDepth$ big enough beyond which nodes are no lower than $\Duration/2$ below the accumulation segment.
  
  All in all, the accumulation set is exactly the targeted segment.
\end{proof}

\section{ASM implementation}
\label{sec:ASMImplementation}
Each augmented meta-signal that will translate into a ray of regular signals has its name written slanted, whereas `regular' ones that translate directly into the SM version in next section, like \SigBorder, are upright.

The initial configuration starts with borders at $-1$ and $+1$ as shown in \RefFig{fig:parameters}.
As mentioned, each sub-tree is bounded by a pair of  \SigBorder signals.
Each time, we have to consider three cases depending on how the node was generated (unary, left or right binary).

\RefFigure{fig:delay_from} shows how a \StepDelay is implemented.
It is initiated by a collision between a main (red, dotted black, or blue) which carries the values of \InputS and \InputT and an auxiliary signal (green or purple).
At this collision, when both \InputS and \InputT are greater than $2$, one augmented signal goes forth (green \SigABounceRSlow, \AGCsigPolice{\textsl{sl}} stands for slow) and back (purple \SigABounceL) to the \SigBorder on the  right.
The amount of time waited is directly proportional to the width of the bounds, and with proper speeds the delay can be made half the width as desired.
These speeds are given in the list of meta-signals in \RefFig{fig:ms:encode}.

\newcommand{\FromLSplit}{\DrawSigABounceRLparam(3 * \BorderVB, \BorderVB)(\MiddleH,\BorderB)[VeryDarkGreen, sloped, above]\DrawSigSplitLLparam(-\BorderVB, \BorderVB)(\MiddleH,\BorderB)[red, sloped,above right] }

\newcommand{\FromDelay}{\DrawSigABounceLLparam(- 3 * \BorderVB, \BorderVB)(\MiddleH,\BorderB)[Purple, sloped, above right]\DrawSigDelayLparam(\MiddleH,\BorderVB)(\MiddleH,\BorderB)[sloped, above]}

\newcommand{\FromRSplit}{\DrawSigABounceLLparam(- 3 * \BorderVB, \BorderVB)(\MiddleH,\BorderB)[Purple, sloped, above right]\DrawSigSplitRLparam(\BorderVB, \BorderVB)(\MiddleH,\BorderB)[blue, sloped,above left] }

\begin{figure}[htb]	
  \centering\scriptsize\SetUnitlength{.012\textwidth}\newcommand{\BorderR}{11}\newcommand{\BorderL}{-\BorderR}\newcommand{\BorderB}{0}\newcommand{\BorderVB}{-3}\newcommand{\BorderU}{\BorderR}\pgfmathsetmacro\MiddleH{(\BorderL + \BorderR) / 2}\pgfmathsetmacro\MiddleV{(2*\BorderU + \BorderB) / 3}\newcommand{\DelaySetting}[1]{\centering\small\ ~\scalebox{.9}{\begin{tikzpicture}
        \path[use as bounding box] (\BorderL-1,\BorderVB-1.25) rectangle (\BorderR+1, \BorderU+.5) ;
        \begin{scope}[very thin,draw=DarkGrey,inner sep=.4mm]
          \draw[<->] (0, \BorderVB - 0.4) -- node[below] {\small 1} +(\BorderR,0) ;
          \draw[<->] (-\BorderR, \BorderVB - 0.4) -- node[below] {\small 1} +(\BorderR,0) ;
          \draw[<->] (\BorderR+1,\BorderB) -- node[right] {\small 1} +(0,\BorderU) ;
        \end{scope}
        \DrawSigBorderLUAbove(\BorderL,\BorderVB)(\BorderL, \BorderU)
        \DrawSigBorderLparam(\BorderR,\BorderVB)(\BorderR, \BorderU)[sloped,above left] #1
        \DrawSigABounceRSlowLparam(\MiddleH,\BorderB)(\BorderR, \MiddleV)[VeryDarkGreen, below, sloped]
        \DrawSigABounceLLparam(\BorderR, \MiddleV)(\MiddleH,\BorderU)[above, sloped, Purple]
        \DrawSigDelaymOLparam(\MiddleH,\BorderB)(\MiddleH,\BorderU)[above, sloped]
      \end{tikzpicture}}~\ }\hspace*{\fill}\subcaptionbox{\StepDelay from left \StepSplit\label{fig:delay_from_LSplit}}{\DelaySetting{\FromLSplit}
  }\hfill\subcaptionbox{\StepDelay from \StepDelay\label{fig:delay_from_delay}}{\DelaySetting{\FromDelay}
  }\hfill\subcaptionbox{\StepDelay from right \mbox{\StepSplit\hspace*{0ex}}\label{fig:delay_from_RSplit}}{\ \DelaySetting{\FromRSplit}\ 
  }\hspace*{\fill}
  \caption{Nodes when $2 \leq \InputS$ and $2 \leq \InputT$.}
  \label{fig:delay_from}
\end{figure}

\begin{figure}[hbt]
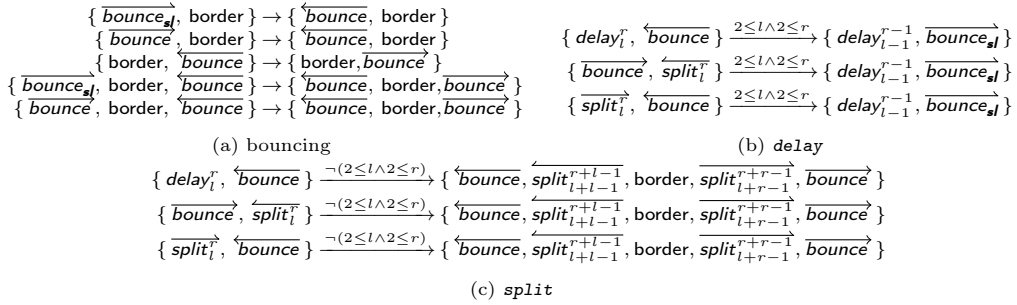

  \centering
  \subcaptionbox{bouncing\label{fig:collision_rules:bouncing}}{\begin{CRlist}
      &\SigABounceRSlow, \SigBorder &\SigABounceL, \SigBorder\\
      &\SigABounceR, \SigBorder &\SigABounceL, \SigBorder\\
      &\SigBorder, \SigABounceL &\SigBorder,\SigABounceR\\
      &\SigABounceRSlow, \SigBorder, \SigABounceL &\SigABounceL, \SigBorder,\SigABounceR\\
      &\SigABounceR, \SigBorder, \SigABounceL &\SigABounceL, \SigBorder,\SigABounceR\\
    \end{CRlist}}\subcaptionbox{\StepDelay\label{fig:collision_rules:delay}}{
    \newcommand{\CRdelayEnd}{$\xrightarrow{2\leq \InputS \land 2 \leq \InputT}$
      &$\SigDelaymO, \SigABounceRSlow$}
    \begin{AugCRlist}
      &\SigDelay, \SigABounceL &\CRdelayEnd\\
      &\SigABounceR, \SigSplitL &\CRdelayEnd\\
      &\SigSplitR, \SigABounceL &\CRdelayEnd\\
    \end{AugCRlist}}
  \subcaptionbox{\StepSplit\label{fig:collision_rules:split}}{
    \newcommand{\CRsplitEnd}{$\xrightarrow{\lnot (2\leq \InputS \land 2 \leq \InputT)}$ &$\SigABounceL, \SigSplitXL, \SigBorder, \SigSplitXR, \SigABounceR$}
    \begin{AugCRlist}
      &\SigDelay, \SigABounceL &\CRsplitEnd\\
      &\SigABounceR, \SigSplitL &\CRsplitEnd\\
      &\SigSplitR, \SigABounceL &\CRsplitEnd\\
    \end{AugCRlist}}
  \caption{Collision rules.}
\end{figure}

Auxiliary signals (\SigABounceR and the likes) bouncing off the borders is described by the collision rules in \RefFig{fig:collision_rules:bouncing}.
The last two rules deal with the case where two such signals bounce at the same point of the border from both sides.

The three rule patterns of \RefFig{fig:collision_rules:delay} correspond to central collision in the three cases in \RefFig{fig:delay_from}.
The left part of the rules corresponds to collisions happening at nodes (with respect to the simulated algorithm) prompted by a previous node accordingly -whether prompted by a \StepDelay, or either side of a \StepSplit.
The right parts of these rule patterns are the same: it is the signals necessary to perform a delay, i.e. to wait a unit of time.

\RefFigure{fig:split_from} shows how a \StepSplit is carried out: once again with green and purple signals bouncing off the walls, this time intercepting split signals half a unit of time later and  in the middle of the new boundaries.
The border created by a \StepSplit ensures both sides of the computation start in the middle of their own borders and are kept separated.
It also effectively scales these subsequent computations down by a factor of two.
The three rule patterns of \RefFig{fig:collision_rules:split} correspond to central collision in the three cases in \RefFig{fig:split_from}.

\begin{figure}[htb] 
  \scriptsize\SetUnitlength{.0155\textwidth}\newcommand{\BorderR}{11}\newcommand{\BorderL}{-\BorderR}\newcommand{\BorderB}{0}\newcommand{\BorderVB}{-2}\pgfmathsetmacro\BorderU{\BorderR * .6}\pgfmathsetmacro\BorderUtop{\BorderR * .75}\pgfmathsetmacro\Width{\BorderR - \BorderL}\pgfmathsetmacro\Height{\BorderU - \BorderB}\pgfmathsetmacro\MiddleH{(\BorderL + \BorderR) / 2}\newcommand{\SplitSetting}[1]{\scalebox{.79}{
      \begin{tikzpicture}
        \path[use as bounding box] (\BorderL-1.75,\BorderVB-1.25) rectangle (\BorderR+1, \BorderUtop) ;
        \begin{scope}[very thin,draw=DarkGrey,inner sep=.4mm]
          \draw[<->] (0, \BorderVB - 0.4) -- node[below] {\small 1} +(\BorderR,0) ;
          \draw[<->] (-\BorderR, \BorderVB - 0.4) -- node[below] {\small 1} +(\BorderR,0) ;
          \draw[<->] (-\BorderR-.5,0) -- node[left] {\small \!1\!/\!2} +(0,\BorderU) ;
        \end{scope}
        \DrawSigBorderLparam(\BorderL, \BorderVB)(\BorderL, \BorderUtop)[pos=0,sloped, below right] \DrawSigBorderLparam(\BorderR, \BorderVB)(\BorderR, \BorderUtop)[sloped, below] \path (\BorderL + \Width / 2,\BorderB) \CoorNode {init} ;
        \path (\BorderL + \Width / 4,\BorderU) node[circle] (left) {} ;
        \path (\BorderR - \Width / 4,\BorderU) node[circle] (right) {} ;
        \DrawSigSplitXLLparam(left)(init)[pos=0, sloped, above right, red]
        \DrawSigSplitXRLparam(init)(right)[pos=1, sloped, above left, blue]
        #1
        \DrawSigBorderLparam(init)(\BorderL + \Width / 2, \BorderUtop)[sloped,pos=1,above left] \DrawSigABounceLLparam(init)(\BorderL, \BorderB + 2 * \Height / 3)[below, sloped, Purple]
        \DrawSigABounceRLparam(\BorderL, \BorderB + 2 * \Height / 3)(left)[above, sloped, VeryDarkGreen]
        \DrawSigABounceLLparam(\BorderR, \BorderB + 2 * \Height / 3)(right)[above, sloped, Purple]
        \DrawSigABounceRLparam(init)(\BorderR, \BorderB + 2 * \Height / 3)[above, sloped, VeryDarkGreen]
      \end{tikzpicture}}}
  \hspace*{\fill}\subcaptionbox{\StepSplit from left \StepSplit\label{fig:split_from_LSplit}}{\SplitSetting{\FromLSplit}
  }\hspace*{\fill}\subcaptionbox{\StepSplit from \StepDelay\label{fig:split_from_delay}}{\SplitSetting{\FromDelay}
  }\hspace*{\fill}\subcaptionbox{\StepSplit from right \StepSplit\label{fig:split_from_RSplit}}{\SplitSetting{\FromRSplit}
  }\hspace*{\fill}\caption{Nodes when $\InputS<2$ or $\InputT<2$.}
  \label{fig:split_from}
\end{figure}

\begin{figure}[hbt]
  \centering
  \begin{tabular}{r@{\,}c@{\,}l}
    \SpeedBounce &=& \SpeedBouncevalue \\[.3em]
    \SpeedBounceSlow &=& \SpeedBounceSlowvalue
  \end{tabular}\quad\begin{MSlist}
    &\SigBorder &0 \\
    &\SigABounceRSlow &\SpeedBounceSlow\\
    &\SigABounceL &-\SpeedBounce\\
    &\SigABounceR &\SpeedBounce\\
  \end{MSlist}
  \qquad
  \begin{MSlist}[Augmented ]
    & $\InputS, \InputT \in [1, +\infty)$ \  \SigDelay &0\\
    & $\InputS, \InputT \in [1, +\infty)$ \  \SigSplitR &1\\
    & $\InputS, \InputT \in [1, +\infty)$ \  \SigSplitL &${-}1$\\
  \end{MSlist}
  \caption{Regular and augmented meta-signals.}
  \label{fig:ms:encode}
\end{figure}

\section{SM implementation}
\label{sec:implementation}

It is first presented how bouncing signals are handled and how augmented signals are encoded as \emph{macro-signals}: coherent packs of parallel signals (also seen as \emph{rays}) that can store the analogue information.
The update and routing is then presented in three stages.
The first stage tests whether the (next) step is \StepDelay or \StepSplit.
The second stage is the routing in a \emph{macro-collision} (coherent pack of collisions issued from a collision with one or more macro-signals, implement an augmented collision).
The third stage is the updating of the parameters.

Please note that the updating of parameters is started at the end of the routing macro-collision and that the test for next step is done right after.
This ensures that the values are updated and the routing scheme is known before the next step starts.

For the sake of readability, signals that are redundant or irrelevant to the construction are often removed from schematic pictures.
Figures without annotation (signal names) are generated through a Java SM simulator and are complete and thorough.

The colour scheme of these pictures is as follows:
\begin{itemizeCompact}
\item border signals are black;
\item tree signals are blue\,---tree signals are those at the exact same position as the augmented signal in the augmented signal machine solution of \RefSec{sec:ASMImplementation};
\item tree macro-signals are filled with yellow;
\item bouncing signals are green, slow bouncing signals are green (and appear almost black in the generated pictures);
\item signals helping with the geometry of macro-collisions are orange;
\item purple, green and light green are used for computation within a macro-signal (in preparation for collision), with purple more often associated to testing and green to changing value;
\item other colours are used for more particular signals and explained in due time;
\item signals foreshadowing or helping to build a \StepSplit are dashed; and
\item signals foreshadowing or helping to build a \StepDelay are dotted.
\end{itemizeCompact}
Although special care has been taken so that pictures remain readable in black and white.

\subsection{ASM simulation structure}
\label{subsec:outsideStructure}

As in \RefSec{sec:ASMImplementation}, the tree structure is built with the help of fast signals bouncing on borders.
The initial configuration thus has two border signals, defining the unit of space.

\begin{figure}[hbt]
  \centering\SetUnitlength{3.5em}\newcommand{\WID}{3}\newcommand{\Width}{.4}\newcommand{\Uorig}{2.75}\newcommand{\TOP}{3}\newcommand{\CoefWaitReturn}{9/4}\begin{tikzpicture}
  \path
  (-\WID,0) \CoorNode{L} +(0,\TOP) \CoorNode{L-t}
  (0,0) \CoorNode{O}  +(0,\TOP) \CoorNode{O-t}
  (\WID,0) \CoorNode{R}  +(0,\TOP) \CoorNode{R-t} 
  (\Width,0) \CoorNode{w0}
  (\Uorig,0) \CoorNode{t0} +(-\CoefWaitReturn*\Width,0)  \CoorNode{b0} 
  ;
  \path[clip,use as bounding box] ([xshift=-.1ex]L) rectangle ([xshift=.1ex]R-t) ;
  \SetIntersectLines[t1]{O}{0}{t0}{-3}
  \SetIntersectLines[tr1]{t1}{3}{R}{0}
  \SetIntersectLines[b1]{O}{0}{b0}{-3}
  \SetIntersectLines[w1]{w0}{0}{t1}{3}
  \SetIntersectLines[br1]{b1}{3}{R}{0}
  \SetIntersectLines[br2]{b1}{1}{br1}{-3}
  \SetIntersectLines[Ol]{-\WID/2,0}{0}{t1}{-1}
  \SetIntersectLines[bl1]{-\WID,0}{0}{b1}{-3}
  \SetIntersectLines[tl1]{-\WID,0}{0}{t1}{-3}
  \SetIntersectLines[tl3]{0,0}{0}{Ol}{3/2}
  \SetIntersectLines[bl2]{-\WID/2,0}{0}{b1}{-1}
  \SetIntersectHeight[Ol-t]{Ol}{0}{\TOP}
  \SetIntersectLines[Or]{\WID/2,0}{0}{t1}{1}
  \SetIntersectLines[Orl]{\WID/4,0}{0}{Or}{-1}
  \SetIntersectLines[Orr]{\WID*3/4,0}{0}{Or}{1}
  \SetIntersectLines[xx-l]{bl2}{3/5}{Ol}{-1}
  \SetIntersectHeight[w-t]{xx-l}{0}{\TOP}
  \SetIntersectLines[trr3]{Or}{3}{R}{0}
  \SetIntersectLines[trl3]{Or}{-3}{O}{0}
  \SetIntersectLines[yy-l]{$(bl2)!.5!(Ol)$}{3/2}{Ol}{-1}
  \SetIntersectLines[bl3]{yy-l}{3/2}{O}{0}
  \path ($(trr3)-.5*(tr1)+.5*(br1)$) \CoorNode{brr3} ;
  \path ($(trl3)-.5*(tr1)+.5*(br1)$) \CoorNode{brl3} ;
  \SetIntersectLines[brl4]{brl3}{3}{Orl}{0}
  \SetIntersectLines[brr4]{brr3}{-3}{Orr}{0}
  \SetIntersectLines[xr]{Or}{-3}{brr3}{3}
  \SetIntersectLines[xl]{Or}{1}{brl3}{-3}
  \SetIntersectLines[wr2]{brr4}{1}{Or}{-3}
  \SetIntersectLines[wl2]{brl4}{-1}{Or}{1}
  \fill[ray_style,thick] (O)
  -- (b1) -- (bl2) -- (Ol-t)
  -- (w-t) -- (xx-l) -- (t1)
  -- (wl2) -- (brl4) -- (Orl) -- (Or)
  -- (Orr) -- (brr4) -- (wr2)
  -- (br2) -- (w1) -- (w0)
  -- cycle ;
  \DrawSigBorder(L)(L-t)
  \DrawSigTree(O)(t1)
  \DrawSigTree(t1)(Ol)
  \DrawSigTree(Ol)(Ol-t)
  \DrawSigTree(t1)(Or)
  \DrawSigTree(Or)(Orl)
  \DrawSigTree(Or)(Orr)
  \DrawSigBorder(t1)(O-t)
  \DrawSigBorder(R)(R-t)
  \DrawSigBorder(bl2)(Ol)
  \DrawSigBounceLTop(t0)(t1)
  \DrawSigBounceLBot(b0)(b1)
  \DrawSigBoundSL(b1)(bl2)
  \DrawSigBounceLBot(b1)(bl1)
  \DrawSigBounceLBot(bl1)(bl2)
  \DrawSigBounceLTop(t1)(tl1)
  \DrawSigBounceRTop(tl1)(Ol)
  \DrawSigBounceRSlowTop(Ol)(tl3)
  \DrawSigBounceRSlowBot(yy-l)(bl3)
  \DrawSigBound(xx-l)(w-t)
  \DrawSigBounceLTop(t1)(tr1)
  \DrawSigBounceRBot(b1)(br1)
  \DrawSigBounceLBot(br1)(br2)
  \DrawSigBounceLTop(tr1)(Or)
  \DrawSigBounceLTop(Or)(trl3)
  \DrawSigBounceRTop(Or)(trr3)
  \DrawSigBounceRTop(trl3)(Orl)
  \DrawSigBounceLTop(trr3)(Orr)
  \DrawSigBound(w0)(w1)
  \DrawSigBound(w1)(br2)
  \DrawSigBound(br2)(wr2)
  \DrawSigBound(wr2)(brr4)
  \DrawSigBound(br2)(wl2)
  \DrawSigBound(wl2)(brl4)
  \DrawSigBounceRBot(bl2)(yy-l)
  \DrawSigBounceRBot(br2)(xr)
  \DrawSigBounceRBot(xr)(brr3)
  \DrawSigBounceRBot(br2)(xl)
  \DrawSigBounceRBot(xl)(brl3)
  \DrawSigBounceRBot(brl3)(brl4)
  \DrawSigBounceRBot(brr3)(brr4)
\end{tikzpicture}
  \caption{Ray scheme example: after a \StepDelay, a \StepSplit then a \StepDelay on the left and a \StepSplit on the right.}
  \label{fig:ray-scheme}
\end{figure}
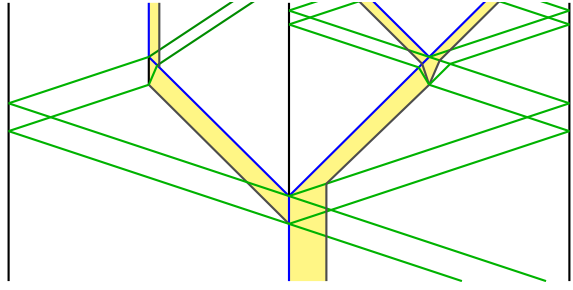

The way the ASM is simulated by a SM is depicted in \RefFig{fig:ray-scheme}.
The filled area corresponds to macro-signals and macro-collisions of the tree.
This area is also used to carry out updating.
The rest are regular signals used for constructing the \StepDelay and \StepSplit steps as previously, but some are doubled to take the width of the macro-signals into account.

As can be seen in \RefFig{fig:ray-scheme}, some meta-signals of the augmented signal machine implementation have to be doubled in order to deal with the width of the macro-signals.
This is the case for \SigABounceRSlow, \SigABounceR and \SigABounceL.
For example, \SigABounceR is replaced by a pair (\SigBounceRTop,\SigBounceRBot) where \SigBounceRTop should be ``above'' \SigBounceRBot.

These fast signals as well as borders as well as relevant collision rules are defined in \RefFig{fig:bounce}.
These definitions are quite straightforward from the ASM definition and are not exemplified.
Their actions can be seen in figures\,\ref{fig:ray-scheme} and \ref{fig:concl:example}.

\begin{figure}[hbt]
  \centering
  \newcommand{\DoList}[1]{\csname SigBounce#1Bot\endcsname,
    \csname SigBounce#1Top\endcsname
  }\begin{MSlist}
    & \SigBorder & 0\\
    & \DoList{RSlow} & \SpeedBounceSlow\\
    & \DoList{R} & \SpeedBounce\\
    & \DoList{L} & -\SpeedBounce\\
  \end{MSlist}
  \qquad
  \begin{CRlist}
    $\forall i\in \{\text{bot}, \text{top}\}$ &\SigBounceRSlowI, \SigBorder&\SigBounceLI, \SigBorder\\
    $\forall i\in \{\text{bot}, \text{top}\}$ &\SigBounceRI, \SigBorder&\SigBounceLI, \SigBorder\\
    $\forall i\in \{\text{bot}, \text{top}\}$ &\SigBounceLI, \SigBorder&\SigBounceRI, \SigBorder\\
  \end{CRlist}
  \\[2mm]
  \begin{CRlist}
    $\forall i, j \in \{\text{bot}, \text{top}\}$  &\SigBounceRSlowI, \SigBorder, \SigBounceLJ&\SigBounceLI, \SigBorder, \SigBounceRJ\\
    $\forall i, j \in \{\text{bot}, \text{top}\}$  &\SigBounceRI, \SigBorder, \SigBounceLJ&\SigBounceLI, \SigBorder, \SigBounceRJ\\
  \end{CRlist} 
  \caption{Definitions for bouncing signals.}
  \label{fig:bounce}
\end{figure}

The height of the bouncing part, \HeightBouncing, and the width of the tree macro-signals, \WidthTree, verify:
$\HeightBouncing=\WidthTree$ on both output of \StepSplit
and
$\HeightBouncing=4/3\WidthTree$ after \StepDelay.
To keep coherence, the following updates are made through steps:
\begin{itemizeCompact}
\item through a \StepSplit from a \StepSplit,
  $\NewHeightBouncing=1/2\HeightBouncing$ and $\NewWidthTree=1/2\WidthTree$,
\item through a \StepSplit from a \StepDelay,
  $\NewHeightBouncing=1/2\HeightBouncing$ and $\NewWidthTree=3/8\WidthTree$,
\item through a \StepDelay from a \StepSplit, 
  $\HeightBouncing=4/3\WidthTree$, and
\item through a \StepDelay from a \StepDelay,
  unchanged.
\end{itemizeCompact}
These remain coherent as long as the initial values are.
Although the computations are not detailed here, these relations are satisfied by all the following constructions and can be checked from the speeds of the signals involved.

\subsection{Encoding of macro-signals}
\label{subsec:Encoding}

\begin{figure}[hbt]
  \SetUnitlength{3.5em}\newcommand{\TOP}{1}\newcommand{\DoPic}[3]{\scalebox{.73}{\begin{tikzpicture}[x={(#1)},y={(#2)}]
        \begin{scope}[inner sep=0.3ex]
          \path (0,0) node [below] {\scriptsize 0} ;
          \path (1,0) node [below] {\scriptsize 1} ;
          \path (2,0) node [below] {\scriptsize 2} ;
          \path (2.8,0) node [below] {\scriptsize \InputS} ;
          \path (2.4,0) node [below] {\scriptsize \InputT} ;    
        \end{scope}
        \fill[ray_style] (0,0)-- (0,\TOP)--(3.8,\TOP)--(3.8,0) --cycle;
        \csname DrawSig#3TreeLUAbove\endcsname(0,0)(0,\TOP)
        \csname DrawSig#3TestLparam\endcsname(.1,0)(.9,\TOP)[pos=.7,sloped,above=-.2em]
        \csname DrawSig#3OneLUBelow\endcsname(1,0)(1,\TOP)
        \csname DrawSig#3TwoLUAbove\endcsname(2,0)(2,\TOP)
        \csname DrawSig#3LLUAbove\endcsname(2.8,0)(2.8,\TOP)
        \csname DrawSig#3RLUAbove\endcsname(2.4,0)(2.4,\TOP)
        \csname DrawSig#3BoundLUAbove\endcsname(3.8,0)(3.8,\TOP)
      \end{tikzpicture}}}\centerline{
    \subcaptionbox{\StepDelay\label{fig:encoding:delay}}{\DoPic{1,0}{0,1}{}
    }
    \quad\subcaptionbox{\label{fig:encoding:split:left}left \StepSplit}{\DoPic{-1,0}{1,1}{SL}
    }\quad\subcaptionbox{\label{fig:encoding:split:right}right \StepSplit}{\DoPic{1,0}{1,1}{SR}
    }}
  \caption{Encoding of augmented signals at leaving a macro-collision.}
  \label{fig:encoding}
\end{figure}
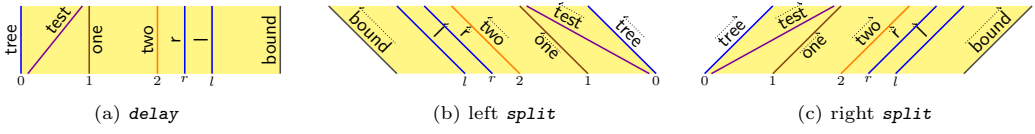

The augmented signals percolating through the tree, \SigDelay, \SigSplitR and \SigSplitL, carry analogue information.
Thus they have to be encoded as by macro-signals with multiple (regular) signals.
The way they are encoded after an updating is depicted in \RefFig{fig:encoding}.
The sequence of parallel signals is to be understood as follows.

The \SigTree (and variants) signals are at the exact locations of the augmented signals in the infinite tree.
Then come signal \SigOne (brown) and signal \SigTwo (orange) which provide the local scale.
Values are encoded by the distance of the signal to \SigTree (so $0$ is at \SigTree).
Signal \SigBound marks the other end of the macro-signal.
These four signals are structural and not affected by the updating of the parameters.

The parameters \InputS and \InputT are encoded by \SigL and \SigR blue signals between \SigOne (included) and \SigBound (excluded).
Thanks to \RefLem{lem:Split_Col}, all the values of the parameters are bounded by $2 + \Slope$ from the start so that the scale can be set to ensure that \SigL and \SigR always remain between \SigOne and \SigBound.
If the value \InputS (or \InputT or both) is equal to $1$ or $2$, then a special signal is used that amount for the superposition.
These special cases are straightforward and are not addressed anymore.

An extra signal with a different speed, \SigTest, initiates the test for the next step inside the macro-signal as presented later on.
Other signals might be present inside the macro-signals to carry out the parameter updating.
The scale is small enough to ensure that all involved signals remain between \SigTree and \SigBound during parameter updates and that any computation finishes before the next macro-collision starts.
Throughout the updating, signals and positions changed, but the encoding remains similar.

Each macro-signal has the same speed as the augmented signal it encodes: 0 for \SigDelay as in \RefFig{fig:encoding:delay}, 1 for \SigSplitR as in \RefFig{fig:encoding:split:right} and -1 for \SigSplitL as in \RefFig{fig:encoding:split:left}.
The sequence of signals is displayed left to right except for \SigSplitL as in \RefFig{fig:encoding:split:left}.

In the rest of this paper, meta-signals exclusively used to form macro-signal \SigDelay, \SigSplitR and \SigSplitL respectively carry no arrow, a right dotted arrow and a left dotted arrow, irrespective of their speeds and directions.
For example \SigOne, \SigSROne and \SigSLOne each encodes the position of scale 1, but in different macro-signals.
Involved meta-signals (except \SigTest signals listed in the next section) are listed in \RefFigure{fig:ms:tree}.

\begin{figure}[hbt]
  \centering\small\newcommand{\DoList}[1]{\csname Sig#1Tree\endcsname,
    \csname Sig#1Bound\endcsname,
    \csname Sig#1One\endcsname,
    \csname Sig#1Two\endcsname,
    \csname Sig#1L\endcsname,
    \csname Sig#1R\endcsname 
  }
  \begin{MSlist}
    & \DoList{SR}& 1 \\
    & \DoList{}& 0 \\
    & \DoList{SL}& -1 \\
  \end{MSlist}
  \caption{Meta-signals for encoding the augmented meta-signal information.}
  \label{fig:ms:tree}
\end{figure}

\subsection{Test}
\label{subsec:test}

\begin{figure}[hbt]
  \small\SetUnitlength{2.45em}\newcommand{\TOP}{2.9}\hspace*{\fill}\subcaptionbox{from \StepDelay\label{fig:testing:delay:fromdelay}}{\scalebox{.85}{\begin{tikzpicture}\path (0.275,0) \CoorNode{testInit} ;
  \path (1,0) \CoorNode{oneInit} ;
  \path (2,0) \CoorNode{twoInit} ;
  \path (3.8,0) \CoorNode{boundInit} ;
  \path (3.8,\TOP) \CoorNode{boundTop} ;
  \SetIntersectLines[interTwo]{testInit}{\SPEEDCOMPUTEvalue*1.25}{twoInit}{0}
  \SetIntersectLines[interBound]{interTwo}{\SPEEDCOMPUTEvalue*1.25}{boundInit}{0}
  \fill[ray_style] (0,0)--(0,\TOP)--(boundTop)--(boundInit)--cycle ;
  \DrawSigTreeLUAboveRight(0,0)(0,\TOP)
  \DrawSigOneLUAboveRight(oneInit)(1,\TOP)
  \DrawSigTwoLUAboveRight(twoInit)(2,\TOP)
  \DrawSigLLUAboveRight(2.8,0)(2.8,\TOP)
  \DrawSigRLUAboveRight(2.4,0)(2.4,\TOP)
  \DrawSigBoundLparam([xshift=-.01ex]boundInit)(interBound)[sloped, pos=0.5, below=-0.2em]
  \DrawSigBoundDELLUBelow(interBound)(boundTop)
  \DrawSigTestLparam(interTwo)(testInit)[sloped, pos=0.77, above right=-0.2em]
  \DrawSigTestDELLUAbove(interTwo)(interBound)
\end{tikzpicture}}}\hfill\subcaptionbox{\label{fig:testing:delay:from left}\StepSplit from left \StepSplit}{\scalebox{.85}{\begin{tikzpicture}\path (-0.6,0) \CoorNode{testInit} ;
  \path (-1,0) \CoorNode{oneInit} ;
  \path (-2,0) \CoorNode{twoInit} ;
  \path (-3.8,0) \CoorNode{boundInit} ;
  \path (-3.8-\TOP,\TOP) \CoorNode{boundTop} ;
  \SetIntersectLines[interTwo]{testInit}{-\SPEEDCOMPUTEvalue*1.25}{twoInit}{-1}
  \SetIntersectLines[interBound]{interTwo}{-\SPEEDCOMPUTEvalue*1.25}{boundInit}{-1}
  \fill[ray_style] (0,0)--(-\TOP,\TOP)--(boundTop)--(boundInit)--cycle ;
  \DrawSigSLTreeLUAboveLeft(0-\TOP,\TOP)(0,0)
  \DrawSigSLOneLUAboveLeft(-1-\TOP,\TOP)(oneInit)
  \DrawSigSLTwoLUAboveLeft(-2-\TOP,\TOP)(twoInit)
  \DrawSigSLLLUAboveLeft(-3.1-\TOP,\TOP)(-3.1,0)
  \DrawSigSLRLUAboveLeft(-2.4-\TOP,\TOP)(-2.4,0)
  \DrawSigSLBoundLUBelow(interBound)(boundInit)
  \DrawSigSLBoundDELLUBelow(interBound)(boundTop)
  \DrawSigSLTestLparam(testInit)(interTwo)[sloped, pos=0.5, above=-0.2em]
  \DrawSigSLTestDELLUAbove(interTwo)(interBound)
\end{tikzpicture}}}\hfill\subcaptionbox{\label{fig:testing:delay:from right}\StepDelay from right \StepSplit}{\scalebox{.85}{\begin{tikzpicture}\path (0.6,0) \CoorNode{testInit} ;
  \path (1,0) \CoorNode{oneInit} ;
  \path (2,0) \CoorNode{twoInit} ;
  \path (3.8,0) \CoorNode{boundInit} ;
  \path (3.8+\TOP,\TOP) \CoorNode{boundTop} ;
  \SetIntersectLines[interTwo]{testInit}{\SPEEDCOMPUTEvalue*1.25}{1.36,0}{1}
  \SetIntersectLines[interBound]{interTwo}{\SPEEDCOMPUTEvalue*1.25}{boundInit}{1}
  \SetIntersectHeight[testStartInit]{testInit}{-\SPEEDCOMPUTEvalue*1.25}{0}
  \fill[ray_style] (0,0)--(\TOP,\TOP)--(boundTop)--(boundInit)--cycle ;
  \DrawSigSRTreeLUAboveRight(0,0)(0+\TOP,\TOP)
  \DrawSigSROneLUAboveRight(oneInit)(1+\TOP,\TOP)
  \DrawSigSRTwoLUAboveRight(twoInit)(2+\TOP,\TOP)
  \DrawSigSRLLUAboveRight(1.36,0)(1.36+\TOP,\TOP)
  \DrawSigSRRLUAboveRight(2.9,0)(2.9+\TOP,\TOP)
  \DrawSigSRBoundLUBelow(boundInit)(interBound)
  \DrawSigSRBoundSPLLUBelow(interBound)(boundTop)
  \DrawSigSRTestLparam(interTwo)(testInit)[sloped, pos=0.5, above=-0.2em]
  \DrawSigSRTestSPLLUAbove(interTwo)(interBound)
\end{tikzpicture}}}\hspace*{\fill}
  \NTVS
  \caption{Testing for the next step.}
  \label{fig:testing}
\end{figure}
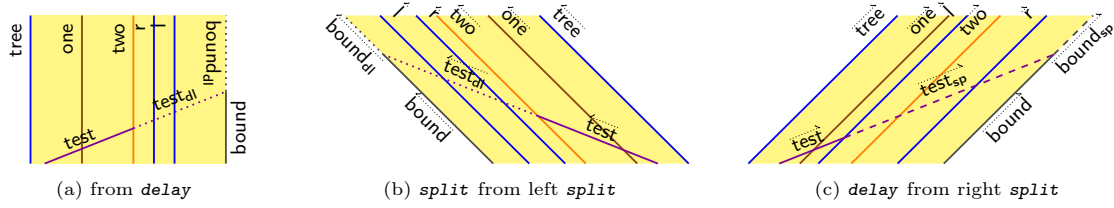

The first stage is to test whether the next step is \StepDelay or \StepSplit, \ie whether \InputS and \InputT are both greater than or equal to $2$.
In order to do so with the encoding, starting from \SigTree, the test is true if \SigTwo is met before any \SigL or \SigR.
So, as shown in \RefFig{fig:testing}, a purple \SigTest signal starts from \SigTree and turns to some signal recording whether the next node ought to be \StepDelay (\SigTestDEL, dotted) or \StepSplit (\SigTestSPL, dashed).
The information is then \emph{stored} on the last signal \SigBound, which becomes \SigBoundDEL (dotted) or \SigBoundSPL (dashed) accordingly.
All signals in the macro-signals are then parallel and nothing happens until the next macro-collision.

Figures\,\ref{fig:testing:delay:fromdelay} and \ref{fig:testing:delay:from left} provide an example where the test lead to \StepDelay while \RefFig{fig:testing:delay:from right} provides one leading to \StepSplit.

\begin{figure}[hbt]
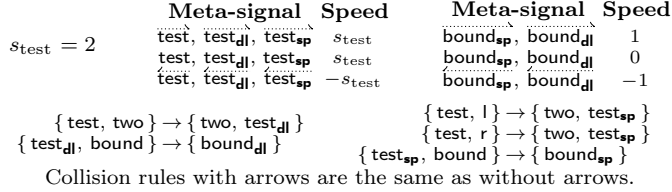

  \small\centering{\footnotesize\SPEEDCOMPUTE$ = \SPEEDCOMPUTEvalue$}
  \qquad\newcommand{\DoList}[1]{\csname Sig#1Test\endcsname,
    \csname Sig#1TestDEL\endcsname,
    \csname Sig#1TestSPL\endcsname
  }\begin{MSlist}
    & \DoList{SR}& \SPEEDCOMPUTE \\
    & \DoList{}& \SPEEDCOMPUTE \\
    & \DoList{SL}& -\SPEEDCOMPUTE 
  \end{MSlist}\qquad\renewcommand{\DoList}[1]{\csname Sig#1BoundSPL\endcsname,
    \csname Sig#1BoundDEL\endcsname
  }\begin{MSlist}
    & \DoList{SR}& 1 \\
    & \DoList{}& 0 \\
    & \DoList{SL}& -1 \\
  \end{MSlist}

  \smallskip
  \begin{CRlist}
    &\SigTest, \SigTwo &\SigTwo, \SigTestDEL\\
    &\SigTestDEL, \SigBound &\SigBoundDEL\\
  \end{CRlist}
  \quad\begin{CRlist}
    &\SigTest, \SigL &\SigTwo, \SigTestSPL\\
    &\SigTest, \SigR &\SigTwo, \SigTestSPL\\
    &\SigTestSPL, \SigBound &\SigBoundSPL\\
  \end{CRlist}
  \centerline{\footnotesize Collision rules with arrows are the same as without arrows.}
  \caption{Definitions for the test.}
  \label{fig:def:test}
\end{figure}

The meta-signals and collision rules at play are listed in \RefFigure{fig:def:test}.

\subsection{Rerouting}

The second stage is the macro-collision started by either \SigBounceLBot or \SigBounceRBot colliding with the macro-signal.
It amounts for the next \emph{tick} of the clock: the time to start generating the next node of the tree.
Duration of steps are handled by bouncing signals so that only the macro-collision is addressed.
The cases of \StepDelay and \StepSplit are considered one after the other.
In each case, the previous node has to be considered.

Rerouting is done using common signal machine tools.
In particular, a \emph{\reflection} is when each signal of a macro-signal bounce off a common signal at a common speed.
A \emph{\refraction} is when each signal of a macro-signal takes a new common speed upon crossing a common signal.
In both cases, parallelism ensures that the proportion between distances between signals remains unchanged.

Since signals \SigOne, \SigTwo, \SigL and \SigR are only rerouted and each in the same fashion, in the schematic pictures, for the sake of clarity, only signal \SigOne is shown.

\subsubsection{\StepDelay rerouting}

At a \StepDelay step, the macro-signal is only straightened if it is not already vertical, and bouncing signals are sent.

\paragraph{\StepDelay rerouting from a \StepDelay}
Since the macro-signal is already vertical, bouncing signals have to be sent back and the updating initiated.
This is done as in \RefFig{fig:delay:from:delay}: upon crossing \SigBoundDEL, the bottom bouncing signal \SigBounceLBot takes note that the next intersection is a \StepDelay, becoming \SigBounceDelayL.
It then simply bounces off \SigTree, becoming the slower \SigBounceRSlowBot.
While bouncing, it also turns \SigTree into \SigTreeDelay, indicating the top bouncing signal to just bounce back.
When leaving, the signal \SigBounceRTop sprouts a signal \SigPreComputeMove which will start the updating (parameter update then test) within the macro-signals on meeting \SigTree.

Signals like \SigOne are unaffected and just pass through the macro-collision.

\begin{figure}[hbt]
  \centering\SetUnitlength{2.5em}\scalebox{.85}{\newcommand{\TOP}{5.7}\newcommand{\BOT}{-1.5}\newcommand{\Wid}{2.5}
\newcommand{\Rmax}{5.75}\begin{tikzpicture}\path (0, \BOT) \CoorNode{wallBot} ;
  \path (0, 0) \CoorNode{bounceBotIntWall} ;
  \path (0, \TOP) \CoorNode{wallTop} ;
  \path (\Wid, \BOT) \CoorNode{boundBot} ;
  \path (.7, \BOT) \CoorNode{oneBot} ;
  \SetIntersectHeight[boundTop]{boundBot}{0}{\TOP};
  \SetIntersectHeight[oneTop]{oneBot}{0}{\TOP};
  \SetIntersectHeight[bounceBotStart]{bounceBotIntWall}{-3}{\BOT};
  \SetIntersectLines[bounceBotIntBound]{bounceBotStart}{-3}{boundBot}{0};
  \SetIntersectLines[bounceBotEnd]{bounceBotIntWall}{3/2}{\Rmax, 0}{0};
  \path (bounceBotIntWall) +(0,4/3*\Wid) \CoorNode{bounceTopIntBound} ;
  \SetIntersectLines[bounceTopStart]{bounceTopIntBound}{-3}{\Rmax, 0}{0};
  \SetIntersectLines[bounceTopIntWall]{bounceTopIntBound}{-3}{wallBot}{0};
  \SetIntersectHeight[bounceTopEnd]{bounceTopIntWall}{3/2}{\TOP}
  \SetIntersectLines[bncTopIntBound]{bounceTopIntWall}{3/2}{boundBot}{0}; 
  \SetIntersectHeight[testStart]{bncTopIntBound}{-3}{\TOP}; 
  \fill[ray_style] (wallBot) -- (wallTop) --  (boundTop)  -- (boundBot) -- cycle ;
  \DrawSigTreeLUAboveLeft(wallBot)(bounceBotIntWall)
  \DrawSigTreeDelayLUAbove(bounceBotIntWall)(bounceTopIntWall)
  \DrawSigTreeLUAboveRight(bounceTopIntWall)(wallTop)
  \DrawSigOneLUAbove(oneBot)(oneTop)
  \DrawSigBoundDELLUAboveLeft(boundBot)(bounceBotIntBound)
  \DrawSigBoundLUAbove(bounceBotIntBound)(boundTop)
  \DrawSigBounceLBotLUAboveRight(bounceBotIntBound)(bounceBotStart)
  \DrawSigBounceDelayLLUAboveLeft(bounceBotIntWall)(bounceBotIntBound)
  \DrawSigBounceRSlowBotLUAboveRight(bounceBotIntWall)(bounceBotEnd)
  \DrawSigBounceLTopLUBelowRight(bounceTopIntWall)(bounceTopStart)
  \DrawSigBounceRSlowTopLUBelowRight(bounceTopIntWall)(bounceTopEnd)
  \DrawSigPreComputeMoveLUBelow(bncTopIntBound)(testStart)
  \end{tikzpicture}}
  \caption{\StepDelay after a \StepDelay.}
  \label{fig:delay:from:delay}
\end{figure}

\paragraph{\StepDelay rerouting from a right branch of a \StepSplit}
If the macro-signal comes from the left, it goes first through a \refraction on \SigWallDelay and then a \refraction on \SigBounceDelayL, as in \RefFig{fig:delay:from:Rward} to ensure the correct scaling of the macro-signal (the two \refraction are more visible on \RefFig{fig:delay:from right}).
The intermediate speed is such that the width of the macro-signal is halved.

Again, information about the next step is carried on the bound side of the macro-signal, by \SigSRBoundDEL.
The output of the starting collision (between \SigBounceLBot and \SigSRTree) contains three signals:
A vertical signal \SigTreeDelay that refracts signal in the band once.
A signal \SigBoundDelaySR that is the diffraction of the bound signal, and still carries the \StepDelay information; that way, \SigBounceLTop can become \SigBounceDelayL and handles the second \refraction.
A signal \SigBounceFSR that, upon intersecting with \SigBounceLTop, can start \SigBounceRSlowBot at the correct place.

\begin{figure}[hbt]
  \centering\SetUnitlength{3.6em}\scalebox{.85}{\newcommand{\delayFromBranchSET}{
  \providecommand{\RMax}{4.5}
  \path (0, 0) \CoorNode{wallBot} ;
  \path (0, 2) \CoorNode{wallMid} ;
  \path (0, 4) \CoorNode{wallTop} ;
  \path (0, 3.2) \CoorNode{oneIntWall};
  \SetIntersectLines[bncSlowStart]{wallMid}{3/2}{wallTop}{-3} \SetIntersectLines[bndStart]{wallBot}{\SPEEDSHRINKvalue}{bncSlowStart}{-3}\SetIntersectLines[oneStart]{oneIntWall}{\SPEEDSHRINKvalue}{bncSlowStart}{-3}\SetIntersectHeight[oneTop]{oneStart}{0}{\TOP};
  \SetIntersectHeight[bndTop]{bndStart}{0}{\TOP};
  \SetIntersectHeight[treeTop]{wallTop}{0}{\TOP};
  \SetIntersectHeight[bncTopTop]{wallTop}{3/2}{\TOP};
  \SetIntersectLines[bncBotTop]{bncSlowStart}{3/2}{\RMax, 0}{0};
  \SetIntersectLines[bncTopIntBound]{wallTop}{3/2}{bndStart}{0}; \SetIntersectHeight[preCompEnd]{bncTopIntBound}{-3}{\TOP};
}
\newcommand{\delayFromBranchDRAW}{
  \DrawSigWallDelayLUAbove(wallBot)(wallTop)
  \DrawSigBounceLTop(bncSlowStart)(bndStart)
  \DrawSigBounceDelayLLUAbove(bndStart)(wallTop)
  \DrawSigBounceFSRLUBelow(wallBot)(bncSlowStart)
  \DrawSigBoundDelaySRLUAbove(wallBot)(bndStart)
  \DrawSigOneSRLUBelow(oneIntWall)(oneStart)
  \DrawSigOneLUBelowRight(oneStart)(oneTop)
  \DrawSigBoundLUBelowLeft(bndStart)(bndTop)
  \DrawSigTreeLUAboveRight(wallTop)(treeTop)
  \DrawSigBounceRSlowBotLUAbove(bncSlowStart)(bncBotTop)
  \DrawSigBounceRSlowTopLUBelowRight(wallTop)(bncTopTop)
  \DrawSigPreComputeMoveLUAbove(bncTopIntBound)(preCompEnd)
}

    \newcommand{\TOP}{5.45}\newcommand{\BOT}{-0.5}\begin{tikzpicture}
  \delayFromBranchSET
  \path (-4 + \BOT, \BOT) \CoorNode{treeInit} ;
  \path (\BOT, \BOT) \CoorNode{boundInit} ;
  \path (- 3 * \BOT, \BOT) \CoorNode{bounceBot} ;
  \SetIntersectLines[bounceTop]{{3 * (4 -\BOT)}, \BOT}{-3}{\RMax, 0}{0};
  \SetIntersectHeight[oneInit]{oneIntWall}{1}{\BOT};
  \fill[ray_style] (boundInit) -- (treeInit) --  (wallTop)  -- (treeTop) -- (bndTop) -- (bndStart) -- (wallBot) -- cycle ;
  \delayFromBranchDRAW
  \DrawSigSRTreeLUAbove(treeInit)(wallTop)
  \DrawSigSROneLUAbove(oneInit)(oneIntWall)
  \DrawSigSRBoundDELLUAbove(boundInit)(wallBot)
  \DrawSigBounceLBotLUAbove(bounceBot)(wallBot)
  \DrawSigBounceLTopLUBelow(bounceTop)(bncSlowStart)
\end{tikzpicture}}
  \caption{\StepDelay after a right \StepSplit.}
  \label{fig:delay:from:Rward}
\end{figure}

\paragraph{\StepDelay rerouting from a left branch of a \StepSplit}
The macro-signal comes from the right, it first undergoes a \reflection on \SigWallDelay, then a \refraction on \SigBounceDelayL, as shown in \RefFig{fig:delay:from:Lward} (it is more visible on \RefFig{fig:delay:from left}).
Once again, the intermediate speed is such that the width of the macro-signal is halved.
So after the first \reflection everything behaves the same way than after the first \refraction when coming from the left.

The reason for using a \reflection is the following: information on the next step is carried on the bound side of the macro-signal (\SigSLBoundDEL) and has to be below (before) \SigSLTree; this is why a leftward macro-signal is a mirror image of a rightward macro-signal, rather than a mere slanting of a vertical macro-signal\,---in other words the bound is on the left.

Without a \SigBounceLTop to handle the \refraction, we need another signal, \SigdsepTwo (orange) coming from somewhere else.
A \SigdsepOne (orange as well) signal is thus launched at the bottom intersection with a default fast speed of $-3$.
It intersects the \SigBounceRTop signal and sprout \SigdsepTwo.
The \SigdsepTwo signal in turn has a finely tuned speed so as to intersect with \SigBounceFSR at the right time and place.
There, a \SigBounceLTop is sprouted as if it came from the right and handles the \refraction.

\begin{figure}[hbt]
  \SetUnitlength{4em}\centering
  \scalebox{.85}{\newcommand{\delayFromBranchSET}{
  \providecommand{\RMax}{4.5}
  \path (0, 0) \CoorNode{wallBot} ;
  \path (0, 2) \CoorNode{wallMid} ;
  \path (0, 4) \CoorNode{wallTop} ;
  \path (0, 3.2) \CoorNode{oneIntWall};
  \SetIntersectLines[bncSlowStart]{wallMid}{3/2}{wallTop}{-3} \SetIntersectLines[bndStart]{wallBot}{\SPEEDSHRINKvalue}{bncSlowStart}{-3}\SetIntersectLines[oneStart]{oneIntWall}{\SPEEDSHRINKvalue}{bncSlowStart}{-3}\SetIntersectHeight[oneTop]{oneStart}{0}{\TOP};
  \SetIntersectHeight[bndTop]{bndStart}{0}{\TOP};
  \SetIntersectHeight[treeTop]{wallTop}{0}{\TOP};
  \SetIntersectHeight[bncTopTop]{wallTop}{3/2}{\TOP};
  \SetIntersectLines[bncBotTop]{bncSlowStart}{3/2}{\RMax, 0}{0};
  \SetIntersectLines[bncTopIntBound]{wallTop}{3/2}{bndStart}{0}; \SetIntersectHeight[preCompEnd]{bncTopIntBound}{-3}{\TOP};
}
\newcommand{\delayFromBranchDRAW}{
  \DrawSigWallDelayLUAbove(wallBot)(wallTop)
  \DrawSigBounceLTop(bncSlowStart)(bndStart)
  \DrawSigBounceDelayLLUAbove(bndStart)(wallTop)
  \DrawSigBounceFSRLUBelow(wallBot)(bncSlowStart)
  \DrawSigBoundDelaySRLUAbove(wallBot)(bndStart)
  \DrawSigOneSRLUBelow(oneIntWall)(oneStart)
  \DrawSigOneLUBelowRight(oneStart)(oneTop)
  \DrawSigBoundLUBelowLeft(bndStart)(bndTop)
  \DrawSigTreeLUAboveRight(wallTop)(treeTop)
  \DrawSigBounceRSlowBotLUAbove(bncSlowStart)(bncBotTop)
  \DrawSigBounceRSlowTopLUBelowRight(wallTop)(bncTopTop)
  \DrawSigPreComputeMoveLUAbove(bncTopIntBound)(preCompEnd)
}

    \begin{tikzpicture}
  \newcommand{\LEF}{-7.5}\newcommand{\TOP}{5.45}\newcommand{\BOT}{-0.5}\newcommand{\Rmax}{4}\begin{pgfinterruptboundingbox}
    \path[clip] (\LEF,\BOT-1) rectangle (\Rmax-.35,\TOP+1) ;
  \end{pgfinterruptboundingbox}
  \delayFromBranchSET
  \path (4 - \BOT, \BOT) \CoorNode{treeInit} ;
  \path (-\BOT, \BOT) \CoorNode{boundInit} ;
  \path (3 * \BOT, \BOT) \CoorNode{bounceBot} ;
  \SetIntersectHeight[oneInit]{oneIntWall}{-1}{\BOT};
  \SetIntersectLines[XXX]{wallBot}{3/5}{wallTop}{-1}
  \fill[ray_style] (treeInit) -- (boundInit) --  (wallBot)  -- (treeTop) -- (bndTop) -- (XXX) -- cycle ;
  \delayFromBranchDRAW
  \DrawSigSLTreeLUAbove(treeInit)(wallTop)
  \DrawSigSLOneLUAbove(oneInit)(oneIntWall)
  \DrawSigSLBoundDELLUAbove(boundInit)(wallBot)
  \DrawSigBounceRBotLUBelow(bounceBot)(wallBot)
  \SetIntersectLines[bounceTop]{wallTop}{3}{\LEF, 0}{0};
  \SetIntersectLines[inter]{wallTop}{3}{wallBot}{-3};
  \DrawSigBounceRTopLUAbove(bounceTop)(inter);
  \DrawSigdsepOneLUAbove(wallBot)(inter);
  \DrawSigdsepTwoLUAbove(inter)(bncSlowStart);
\end{tikzpicture}}

  \caption{\StepDelay after a left \StepSplit.}
  \label{fig:delay:from:Lward}
\end{figure}

The new meta-signals and collision rules involved are listed in \RefFigure{fig:def:delay:rerouting}.

\begin{figure}[hbt]
  \centering
  \footnotesize
  \scalebox{1}{
    \begin{tabular}{r@{\,}c@{\,}l}
      \SPEEDSHRINK &=& \SPEEDSHRINKvalue\\[.3em]
      \SpeedBounceFASTSHRINK &=& \SpeedBounceFASTSHRINKvalue
    \end{tabular}\quad\begin{MSlist}
      & \SigTreeDelay, \SigWallDelay &0\\
      & \SigdsepOne & -\SpeedBounce\\
      & \SigdsepTwo & 2\SpeedBounce
    \end{MSlist}\quad\begin{MSlist}
      &\SigBounceDelayL, \SigPreComputeMove & -\SpeedBounce\\&\SigBounceFSR &\SpeedBounceFASTSHRINK\\
      &\SigBoundDelaySR, \SigOneSR, \SigTwoSR, \SigRSR, \SigLSR &\SPEEDSHRINK\\
    \end{MSlist}}
  
  \medskip

  \begin{CRlist}
    &\SigBoundDEL, \SigBounceLBot &\SigBounceDelayL, \SigBound\\
    &\SigBounceDelayL, \SigTree &\SigTreeDelay, \SigBounceRSlowBot\\
    &\SigTreeDelay, \SigBounceLTop &\SigTree, \SigBounceRSlowTop\\
    &\SigSRBoundDEL, \SigBounceLBot &\SigWallDelay, \SigBoundDelaySR, \SigBounceFSR\\
    &\SigBounceFSR, \SigBounceLTop &\SigBounceLTop, \SigBounceRSlowBot\\
    &\SigSRTree, \SigWallDelay, \SigBounceDelayL &\SigTree, \SigBounceRSlowTop\\
    &\SigBounceRBot, \SigSLBoundDEL &\SigdsepOne, \SigWallDelay, \SigBoundDelaySR, \SigBounceFSR\\
    &\SigBounceRTop, \SigdsepOne &\SigdsepTwo\\
    &\SigBounceFSR, \SigdsepTwo &\SigBounceLTop, \SigBounceRSlowBot\\
    &\SigWallDelay, \SigSLTree, \SigBounceDelayL &\SigTree, \SigBounceRSlowTop\\
    &\SigBounceRSlowTop, \SigBound &\SigPreComputeMove, \SigBound, \SigBounceRSlowTop\\[.3em]
    $\forall \AGCmetaSignal \in \{\SigOne, \SigTwo, \SigR, \SigL\}$& \AGCRightWard{\AGCmetaSignal}, \SigWallDelay & \SigWallDelay, \AGCRightWardSlow{\AGCsigPolice{{\AGCmetaSignal}\IndexS}} \\
    $\forall \AGCmetaSignal \in \{\SigOne, \SigTwo, \SigR, \SigL\}$& \AGCLeftWard{\AGCmetaSignal}, \SigWallDelay & \SigWallDelay, \AGCRightWardSlow{\AGCsigPolice{{\AGCmetaSignal}\IndexS}}\\[.3em]
    &\SigBoundDelaySR, \SigBounceLTop &\SigBounceDelayL, \SigBound\\
    $\forall \AGCmetaSignal \in \{\SigOne, \SigTwo, \SigR, \SigL\}$&\AGCRightWardSlow{\AGCsigPolice{{\AGCmetaSignal}\IndexS}}, \SigBounceDelayL & \SigBounceDelayL, \AGCmetaSignal \\
  \end{CRlist}
  \caption{Definitions for rerouting at a \StepDelay node.}
  \label{fig:def:delay:rerouting}
\end{figure}

\subsubsection{Split Rerouting}

At a \StepSplit step we need to fork the macro-signal, and go half a unit of space both ways during half a unit of time (again using bouncing signals).
Forking is done by sending a copy of each side (with corresponding orders of signals inside the new macro-signals) and adding a motionless \SigBorder signal, so as to set border and scale for the new branches.

\paragraph{\StepSplit rerouting from a \StepSplit}
Routing from each branch is done symmetrically (as can be seen in Figs.\,\ref{fig:split:from left} and \ref{fig:split:from right}) so only one case is presented.
Forking the macro-signal from the right branch of a \StepSplit is done by raising a vertical signal \SigWallSplit on which the macro-signal gets duplicated as can be seen in \RefFig{fig:split:from:Rward}.
The duplication can be viewed as a simultaneously operating a \reflection and a \refraction on the macro-signal.
Both branches are then routed and their widths are halved in the process.

\begin{figure}[hbt]
  \centering\SetUnitlength{4.5em}\scalebox{.85}{
\begin{tikzpicture}
  \newcommand{\RITE}{4.5}\newcommand{\LEF}{-5}\newcommand{\TOP}{5.45}\newcommand{\BOT}{-0.5}\path (0, 0) \CoorNode{wallBot} ;
  \path (0, 2) \CoorNode{wallMid} ;
  \path (0, 4) \CoorNode{wallTop} ;
  \path (0, 3.2) \CoorNode{oneIntWall};
  \SetIntersectHeight[treeInit]{wallTop}{1}{\BOT};
  \SetIntersectHeight[boundInit]{wallBot}{1}{\BOT};
  \SetIntersectHeight[bounceBot]{wallBot}{-3}{\BOT};
  \SetIntersectLines[bncStartR]{wallMid}{3}{wallTop}{-3} \SetIntersectLines[bncStartL]{wallMid}{-3}{wallTop}{1} \SetIntersectLines[bndStartR]{wallBot}{\SPEEDSHRINKvalue}{wallTop}{-3}\SetIntersectLines[bndStartL]{wallBot}{-\SPEEDSHRINKBACKvalue}{wallTop}{1}\SetIntersectLines[oneStartR]{oneIntWall}{\SPEEDSHRINKvalue}{wallTop}{-3}\SetIntersectLines[oneStartL]{oneIntWall}{-\SPEEDSHRINKBACKvalue}{wallTop}{1}\SetIntersectHeight[oneInit]{oneIntWall}{1}{\BOT};
  \SetIntersectLines[bounceLTopBot]{wallTop}{-3}{\RITE, 0}{0};
  \SetIntersectHeight[bounceLTopTop]{wallTop}{-3}{\TOP};
  \SetIntersectLines[bounceLBotTop]{bncStartL}{-3}{\LEF, 0}{0};
  \SetIntersectHeight[bounceRTopTop]{wallTop}{3}{\TOP};
  \SetIntersectLines[bounceRBotTop]{bncStartR}{3}{\RITE, 0}{0};
  \SetIntersectHeight[oneRTop]{oneStartR}{1}{\TOP};
  \SetIntersectHeight[oneLTop]{oneStartL}{-1}{\TOP};
  \SetIntersectHeight[bndRTop]{bndStartR}{1}{\TOP};
  \SetIntersectHeight[bndLTop]{bndStartL}{-1}{\TOP};
  \SetIntersectHeight[treeRTop]{wallTop}{1}{\TOP};
  \SetIntersectHeight[treeLTop]{wallTop}{-1}{\TOP};
  \SetIntersectHeight[BorderTop]{wallTop}{0}{\TOP}
  \SetIntersectLines[bncTopIntBoundR]{wallTop}{3}{bndStartR}{1}; 
  \SetIntersectHeight[updtStartR]{bncTopIntBoundR}{-3}{\TOP}; 
  \SetIntersectLines[bncTopIntBoundL]{wallTop}{-3}{bndStartL}{-1}; 
  \SetIntersectHeight[updtStartL]{bncTopIntBoundL}{3}{\TOP}; 
  \path[ray_style] (treeInit) -- (bndStartL) -- (bndLTop) -- (treeLTop) -- (wallTop)  -- (treeRTop) -- (bndRTop) -- (bndStartR) -- (wallBot) -- (boundInit) -- cycle ;  \DrawSigSRTreeLUAbove(treeInit)(wallTop)
  \DrawSigSROneLUAbove(oneInit)(oneIntWall)
  \DrawSigSRBoundSPLLUAbove(boundInit)(wallBot)
  \DrawSigBounceLBotLUAbove(bounceBot)(wallBot)
  \DrawSigWallSplitLUAbove(wallBot)(wallTop)
  \DrawSigBounceSRLUBelow(wallBot)(bncStartR)
  \DrawSigBounceSBLLUBelow(wallBot)(bncStartL)
  \DrawSigBoundSRLUAbove(wallBot)(bndStartR)
  \DrawSigBoundSBLLUAbove(wallBot)(bndStartL)
  \DrawSigOneSRLUBelow(oneIntWall)(oneStartR)
  \DrawSigOneSBLLUBelow(oneIntWall)(oneStartL)
  \DrawSigBounceLTopLparam(wallTop)(bounceLTopBot)[pos=.8,sloped,below];
  \DrawSigBounceLTopLUBelowLeft(bounceLTopTop)(wallTop);
  \DrawSigBounceLBotLUAbove(bncStartL)(bounceLBotTop);
  \DrawSigBounceRTopLUBelowRight(wallTop)(bounceRTopTop);
  \DrawSigBounceRBotLUAbove(bncStartR)(bounceRBotTop);
  \DrawSigSROneLUBelow(oneStartR)(oneRTop)
  \DrawSigSLOneLUBelow(oneLTop)(oneStartL)
  \DrawSigSRBoundLUAboveLeft(bndStartR)(bndRTop)
  \DrawSigSLBoundLUAboveRight(bndLTop)(bndStartL)
  \DrawSigSRTreeLUAboveRight(wallTop)(treeRTop)
  \DrawSigSLTreeLUAboveLeft(treeLTop)(wallTop)
  \DrawSigPreSplitUpdtRLUBelow(updtStartR)(bncTopIntBoundR);
  \DrawSigPreSplitUpdtLLUBelow(bncTopIntBoundL)(updtStartL);
  \DrawSigBorderLUAbove(wallTop)(BorderTop)
  \end{tikzpicture}}\NTVS
  \caption{\StepSplit after a right \StepSplit (routing).}
  \label{fig:split:from:Rward}
\end{figure}

The meta-signals and collision rules at play are listed in \RefFig{fig:def:split:from:split}.

\begin{figure}[hbt]
  \centering{\scriptsize
    \begin{tabular}{r@{\,}c@{\,}l}
      \SPEEDSHRINKBACK &=& \SPEEDSHRINKBACKvalue\\[.3em]
      \SpeedBounceSHRINK &=& \SpeedBounceSHRINKvalue\\[.3em]
      \SpeedBounceSHRINKBACK &=& \SpeedBounceSHRINKBACKvalue\\
    \end{tabular}}\hfill
  \begin{MSlist}
    &\SigWallSplit & 0\\
    &\SigBounceSR & \SpeedBounceSHRINK\\
    &\SigBounceSL & -\SpeedBounceSHRINK\\
    &\SigBounceSBR & \SpeedBounceSHRINKBACK\\
    &\SigBounceSBL & -\SpeedBounceSHRINKBACK\\
  \end{MSlist}\hfill
  \begin{MSlist}
    &\SigBoundSR &\SPEEDSHRINK\\
    &\SigOneSL, \SigTwoSL, \SigRSL, \SigLSL, \SigBoundSL &-\SPEEDSHRINK\\
    &\SigOneSBR,\SigTwoSBR, \SigRSBR, \SigLSBR, \SigBoundSBR &\SPEEDSHRINKBACK\\
    &\SigOneSBL, \SigTwoSBL, \SigRSBL, \SigLSBL,\SigBoundSBL &-\SPEEDSHRINKBACK\\
    &\SigPreSplitUpdtR &-\SPEEDCOMPUTE\\
    &\SigPreSplitUpdtL &\SPEEDCOMPUTE
  \end{MSlist}
  \medskip
  \small
  \begin{tabular}{c}
    \begin{CRlist}
      &\SigBounceLBot, \SigSRBoundSPL &\SigBounceSBL, \SigBoundSBL, \SigWallSplit, \SigBoundSR, \SigBounceSR\\
      &\SigBounceRBot, \SigSLBoundSPL &\SigBounceSBR, \SigBoundSBR, \SigWallSplit, \SigBoundSL, \SigBounceSL\\
      &\SigSRTree, \SigWallSplit, \SigBounceLTop &\SigBounceLTop, \SigSplitUpdtLowL, \SigSplitUpdtHighL, \SigSLTree, \SigBorder, \SigSRTree, \SigBounceRTop
    \end{CRlist}
    \\[.3em]
    \begin{CRlist}
      &\SigBoundSR, \SigBounceLTop &\SigBounceLTop, \SigSRBound\\
      &\SigSRTree, \SigBoundSBL &\SigSLBound, \SigSRTree\\
      $\forall \AGCmetaSignal \in \{\SigOne, \SigTwo, \SigR, \SigL\}$&\SigWallSplit, \AGCRightWard{\AGCmetaSignal} &\AGCLeftWardSlow{\AGCsigPolice{{\AGCmetaSignal}\IndexSB}}, \SigWallSplit, \AGCRightWardSlow{\AGCsigPolice{{\AGCmetaSignal}\IndexS}}\\
      $\forall \AGCmetaSignal \in \{\SigOne, \SigTwo, \SigR, \SigL\}$&\AGCRightWardSlow{\AGCsigPolice{{\AGCmetaSignal}\IndexS}}, \SigBounceLTop &\SigBounceLTop, \AGCRightWard{\AGCmetaSignal}\\
      $\forall \AGCmetaSignal \in \{\SigOne, \SigTwo, \SigR, \SigL\}$&\SigSRTree, \AGCLeftWardSlow{\AGCsigPolice{{\AGCmetaSignal}\IndexSB}} &\AGCLeftWard{\AGCmetaSignal}, \SigSRTree\\
      &\SigSRTree, \SigBounceSBL &\SigBounceLBot, \SigSRTree\\
      &\SigBounceRTop, \SigSRBound &\SigPreSplitUpdtR, \SigBounceRTop, \SigSRBound
      \\[.6em]
      &\SigBoundSL, \SigBounceRTop &\SigBounceRTop, \SigSLBound\\
      &\SigSLTree, \SigBoundSBR &\SigSRBound, \SigSLTree\\
      $\forall \AGCmetaSignal \in \{\SigOne, \SigTwo, \SigR, \SigL\}$&\SigWallSplit, \AGCLeftWard{\AGCmetaSignal} &\AGCLeftWardSlow{\AGCsigPolice{{\AGCmetaSignal}\IndexS}}, \SigWallSplit, \AGCRightWardSlow{\AGCsigPolice{{\AGCmetaSignal}\IndexSB}}\\
      $\forall \AGCmetaSignal \in \{\SigOne, \SigTwo, \SigR, \SigL\}$&\AGCLeftWardSlow{\AGCsigPolice{{\AGCmetaSignal}\IndexS}}, \SigBounceRTop &\SigBounceRTop, \AGCLeftWard{\AGCmetaSignal}\\
      $\forall \AGCmetaSignal \in \{\SigOne, \SigTwo, \SigR, \SigL\}$&\SigSLTree, \AGCRightWardSlow{\AGCsigPolice{{\AGCmetaSignal}\IndexSB}} &\AGCRightWard{\AGCmetaSignal}, \SigSLTree\\
      &\SigSLTree, \SigBounceSBR &\SigBounceRBot, \SigSLTree\\
      &\SigBounceLTop, \SigSLBound &\SigPreSplitUpdtL, \SigBounceLTop, \SigSLBound\\
      \end{CRlist}\end{tabular}
  \caption{Definitions for routing at a \StepSplit from right or left.}
  \label{fig:def:split:from:split}
\end{figure}

\paragraph{\StepSplit rerouting from a \StepDelay}
If the previous step is a \StepDelay, the construction, shown in \RefFig{fig:split:from:Delay}, is more involving.
Again this is due to the difference of orientation between the signals in a leftward band and a straight one.
The right part of the collision output is obtained by a simple \refraction on signal \SigBounceSplitOneL.
The left part is obtained through a \refraction by \SigsepOne, a \reflection on \SigBoundSOne and a \refraction on \SigsepThree.

\begin{figure}[hbt]
  \centering\SetUnitlength{2.4em}\scalebox{.85}{\newcommand{\TOP}{6.5}\newcommand{\BOT}{-2}\newcommand{\RIGHT}{6.5}\newcommand{\LEFT}{-6}\begin{tikzpicture}
  \path (0, 0) \CoorNode{wallBot} ;
  \SetIntersectHeight[treeInit]{wallBot}{0}{\BOT}  
  \path (.75,\BOT) \CoorNode{oneInit} ;
  \path (1.5, \BOT) \CoorNode{twoInit} ;
  \path (3.8, \BOT) \CoorNode{boundInit} ;
  \SetIntersectLines[bounceBotIntBound]{wallBot}{-3}{boundInit}{0} ;
  \SetIntersectHeight[bounceBotInit]{wallBot}{-3}{\BOT} ;
  \SetIntersectLines[sep1IntBound]{wallBot}{3}{boundInit}{0} ;
  \SetIntersectLines[bndStartL]{sep1IntBound}{-3}{wallBot}{-1} ;
  \SetIntersectLines[wallTop]{bndStartL}{3}{wallBot}{0} ;
  \SetIntersectLines[bndStartR]{wallTop}{-3}{boundInit}{0} ;
  \SetIntersectLines[bounceTopInit]{wallTop}{-3}{\RIGHT, \BOT}{0} ;
  \SetIntersectLines[oneStartR]{wallTop}{-3}{oneInit}{0} ;
  \SetIntersectLines[twoStartR]{wallTop}{-3}{twoInit}{0} ;
  \SetIntersectLines[oneIntSep1]{wallBot}{3}{oneInit}{0} ;
  \SetIntersectLines[twoIntSep1]{wallBot}{3}{twoInit}{0} ;
  \SetIntersectLines[boundIntSep1]{wallBot}{3}{boundInit}{0} ;
  \SetIntersectLines[oneS1IntBound]{wallBot}{-1}{oneIntSep1}{-3} ;
  \SetIntersectLines[twoS1IntBound]{wallBot}{-1}{twoIntSep1}{-3} ;
  \SetIntersectLines[oneStartL]{oneS1IntBound}{0}{wallTop}{3} ;
  \SetIntersectLines[twoStartL]{twoS1IntBound}{0}{wallTop}{3} ;
  \SetIntersectHeight[bounceLTopTop]{wallTop}{-3}{\TOP} ;
  \SetIntersectLines[bounceLBotTop]{wallBot}{-3}{\LEFT, 0}{0};
  \SetIntersectHeight[bounceRTopTop]{wallTop}{3}{\TOP} ;
  \SetIntersectLines[bounceRBotTop]{wallBot}{3}{\RIGHT, 0}{0};
  \SetIntersectHeight[oneRTop]{oneStartR}{1}{\TOP};
  \SetIntersectHeight[oneLTop]{oneStartL}{-1}{\TOP};
  \SetIntersectHeight[twoRTop]{twoStartR}{1}{\TOP};
  \SetIntersectHeight[twoLTop]{twoStartL}{-1}{\TOP};
  \SetIntersectHeight[bndRTop]{bndStartR}{1}{\TOP};
  \SetIntersectLines[bndLTop]{bndStartL}{-1}{\LEFT, 0}{0};
  \SetIntersectHeight[treeRTop]{wallTop}{1}{\TOP};
  \SetIntersectHeight[treeLTop]{wallTop}{-1}{\TOP};
  \SetIntersectHeight[BorderTop]{wallTop}{0}{\TOP}
  \path[ray_style] (treeInit) -- (wallBot) -- (bndLTop) -- (\LEFT,\TOP)
  -- (treeLTop) -- (wallTop) -- (treeRTop) -- (bndRTop) --(bndStartR) -- (boundInit)
  -- cycle ;
  \DrawSigBounceLBotLUAbove(bounceBotInit)(bounceBotIntBound);
  \DrawSigBounceSplitLLUAboveRight(wallBot)(bounceBotIntBound) ;
  \DrawSigsepOneLparam(wallBot)(sep1IntBound)[sloped,below,pos=.7] ;
  \DrawSigBoundSOneLUBelow(wallBot)(bndStartL) ;
  \DrawSigsepTwoLparam(bndStartL)(sep1IntBound)[sloped,above,pos=.7] ;
  \DrawSigsepThreeLUAboveLeft(bndStartL)(wallTop) ;
  \DrawSigTreeLUAboveLeft(treeInit)(wallTop) ;
  \DrawSigBounceLTopLUAbove(bounceTopInit)(bndStartR);
  \DrawSigBounceSplitOneLLUAboveRight(wallTop)(bndStartR) ;
  \DrawSigBoundSPLLLeft(boundInit)(bounceBotIntBound) ;
  \DrawSigBoundSPLOneLUAboveRight(bounceBotIntBound)(bndStartR) ;
  \DrawSigOneLUAboveLeft(oneInit)(oneStartR) ;
  \DrawSigTwoLUAboveLeft(twoInit)(twoStartR) ;
  \DrawSigOneSOneLUAboveRight(oneS1IntBound)(oneIntSep1) ;
  \DrawSigTwoSOneLUAboveRight(twoS1IntBound)(twoIntSep1) ;
  \DrawSigOneLUAboveRight(oneS1IntBound)(oneStartL) ;
  \DrawSigTwoLUAboveRight(twoS1IntBound)(twoStartL) ;
  \DrawSigBounceLTopLUBelowRight(bounceLTopTop)(wallTop) ;
  \DrawSigBounceLBotLUBelowLeft(bounceLBotTop)(wallBot) ;
  \DrawSigBounceRTopLUBelowLeft(wallTop)(bounceRTopTop);  \DrawSigBounceRBotLUBelowRight(sep1IntBound)(bounceRBotTop) ;
  \DrawSigSROneLUBelowRight(oneStartR)(oneRTop)
  \DrawSigSLOneLUBelowLeft(oneLTop)(oneStartL)
  \DrawSigSRTwoLUBelowRight(twoStartR)(twoRTop)
  \DrawSigSLTwoLUBelowLeft(twoLTop)(twoStartL)
  \DrawSigSRBoundLUBelow(bndStartR)(bndRTop)
  \DrawSigSLBoundLUBelow(bndStartL)(bndLTop)
  \DrawSigSRTreeLUBelowRight(wallTop)(treeRTop)
  \DrawSigSLTreeLUBelowLeft(treeLTop)(wallTop)
  \DrawSigBorderLUAboveRight(wallTop)(BorderTop)
\end{tikzpicture}}\caption{Routing for \StepSplit after a \StepDelay.}
  \label{fig:split:from:Delay}
\end{figure}

The new meta-signals and collision rules at play are listed in \RefFig{fig:def:split:from:Delay}.

\begin{figure}[hbt]
  \centering\begin{MSlist}
    & \SigBoundSOne & -1\\
    & \SigsepOne, \SigsepThree & \SpeedBounce\\
    & \SigsepTwo & -\SpeedBounce\\
    & \SigBounceSplitL, \SigBounceSplitOneL & - \SPEEDCOMPUTE\\
  \end{MSlist}
  \medskip
  \begin{tabular}{c}
    \begin{CRlist}
      &\SigBoundSPL, \SigBounceLBot &\SigBounceSplitL, \SigBoundSPLOne\\
      &\SigBoundSPLOne, \SigBounceLTop &\SigBounceSplitOneL, \SigSRBound\\
      &\SigTree,\SigBounceSplitL&\SigBounceLBot,\SigBoundSOne,\SigTree,\SigsepOne\\
      $\forall \AGCmetaSignal \in \{\SigOne, \SigTwo, \SigR, \SigL\}$&\SigsepOne, \AGCmetaSignal &{\AGCmetaSignal}\IndexOne, \AGCmetaSignal, \SigsepOne\\
      &\SigsepOne, \SigBoundSPLOne &\SigsepTwo, \SigBoundSPLOne, \SigBounceRBot\\
      $\forall \AGCmetaSignal \in \{\SigOne, \SigTwo, \SigR, \SigL\}$&\SigBoundSOne, {\AGCmetaSignal}\IndexOne & \SigBoundSOne, \AGCmetaSignal\\
      &\SigBoundSOne, \SigsepTwo &\SigSLBound , \SigsepThree\\
      $\forall \AGCmetaSignal \in \{\SigOne, \SigTwo, \SigR, \SigL\}$&\AGCmetaSignal, \SigBounceSplitOneL & \SigBounceSplitOneL, \AGCRightWard{\AGCmetaSignal}\\
      $\forall \AGCmetaSignal \in \{\SigOne, \SigTwo, \SigR, \SigL\}$&\SigsepThree, \AGCmetaSignal &\SigsepThree, \AGCLeftWard{\AGCmetaSignal}\\
    \end{CRlist}
    \\
    \begin{CRlist}
      &\SigsepThree, \SigTree, \SigBounceSplitOneL &\SigBounceLTop, \SigSLTree, \SigBorder, \SigSRTree, \SigBounceRTop\\
    \end{CRlist}\end{tabular}
  \caption{Definitions for \StepSplit routing after a \StepDelay.}
  \label{fig:def:split:from:Delay}
\end{figure}

\subsection{Updating the parameters}
\label{subsec:updating}

The updating of parameters is done according to \RefAlgo{code:aug-rule}.
It occurs right after a macro-collision and depends on its nature.

\subsubsection{\StepDelay parameter update}

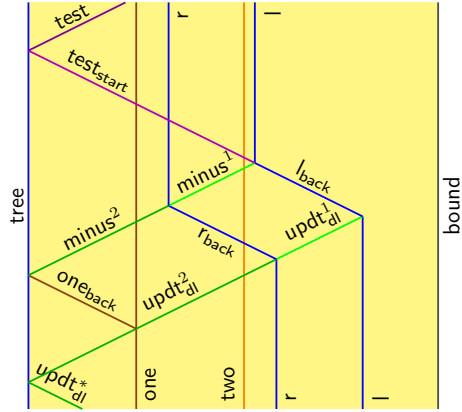
\begin{figure}[hbt]
  \centering\SetUnitlength{4.75em}\scalebox{.85}{\newcommand{\TOP}{3.8}\begin{tikzpicture}
  \path (0,0) \CoorNode{treeInit} ;
  \path (.5,0) \CoorNode{delayStarInit} ;
  \path (1,0) \CoorNode{oneInit} ;
  \path (2,0) \CoorNode{twoInit} ;
  \path (3.8,0) \CoorNode{boundInit} ;
  \path (3.8,\TOP) \CoorNode{boundTop} ;
  \path (3.1,0) \CoorNode{lInit} ;
  \path (2.3,0) \CoorNode{rInit} ;
  \SetIntersectLines[delayInit]{delayStarInit}{-\SPEEDCOMPUTEvalue}{treeInit}{0}
  \SetIntersectLines[delayIntR]{delayInit}{\SPEEDCOMPUTEvalue}{rInit}{0} 
  \SetIntersectLines[delayIntL]{delayInit}{\SPEEDCOMPUTEvalue}{lInit}{0}
  \path[ray_style] (boundTop) -- (boundInit) -- (treeInit) -- (0,\TOP) -- cycle ;
  \DrawSigTreeLUAbove(treeInit)(0,\TOP)
  \DrawSigOneLUBelowLeft(oneInit)(1,\TOP)
  \DrawSigTwoLUAboveLeft(twoInit)(2,\TOP)
  \DrawSigBoundLUBelow(boundInit)(boundTop)
  \DrawSigRLUBelowLeft(rInit)(delayIntR)
  \DrawSigLLUBelowLeft(lInit)(delayIntL)
  \DrawSigPreComputeMoveLUAbove(delayStarInit)(delayInit)
  \DrawSigComputeMoveLparam(delayInit)(delayIntR)[sloped,above=-.2em,pos=.6] 
  \DrawSigComputeMoveOneLUAbove(delayIntR)(delayIntL)
  \SetIntersectLines[delayIntOne]{delayInit}{\SPEEDCOMPUTEvalue}{oneInit}{0}
  \SetIntersectLines[OneBIntT]{delayIntOne}{-\SPEEDCOMPUTEvalue}{treeInit}{0}
  \DrawSigOneBackLUAbove(delayIntOne)(OneBIntT)
  \SetIntersectLines[minusIntR]{OneBIntT}{\SPEEDCOMPUTEvalue}{delayIntR}{-\SPEEDCOMPUTEvalue}
  \SetIntersectLines[minusIntL]{OneBIntT}{\SPEEDCOMPUTEvalue}{delayIntL}{-\SPEEDCOMPUTEvalue}
  \DrawSigRBackLUBelow(delayIntR)(minusIntR)
  \DrawSigLBackLUAbove(delayIntL)(minusIntL)
  \DrawSigMinusLUAbove(OneBIntT)(minusIntR)
  \DrawSigMinusOneLUAbove(minusIntR)(minusIntL)
  \path (minusIntR) ; \pgfgetlastxy{\XCoordR}{\YCoordR} ; \path (minusIntL) ; \pgfgetlastxy{\XCoordL}{\YCoordL} ; \DrawSigRLUBelowRight(minusIntR)(\XCoordR, \TOP)
  \DrawSigLLUBelowRight(minusIntL)(\XCoordL, \TOP)
  \SetIntersectLines[testIntT]{minusIntL}{-\SPEEDCOMPUTEvalue}{treeInit}{0}
  \DrawSigTestStartLparam(minusIntL)(testIntT)[sloped,above=-.2em,pos=.7]
  \SetIntersectHeight[testIntTwo]{testIntT}{\SPEEDCOMPUTEvalue}{\TOP}
  \DrawSigTestLUAbove(testIntT)(testIntTwo)
\end{tikzpicture}}
  \caption{Updating \InputS and \InputT after a \StepDelay.}
  \label{fig:update:delay}
\end{figure}

The updating is done by removing $1$ from both \InputS and \InputT as depicted in \RefFig{fig:update:delay}.
This corresponds to a shift of \SigL and \SigR by the distance from \SigTree to \SigOne.
For \SigL, this is done by constructing a parallelogram with one side going from \SigOne to \SigTree and the opposite side from \SigL to its updated position.
Parallel lines simply correspond to signals with the same speed (\SigOneBack and \SigRBack, and \SigComputeMove and \SigMinus).
Signal \SigR is shifted similarly.
After that, the signal \SigTestStart is generated and sent to collide \SigTree to generate \SigTest to start the test sequence as seen previously.

The new meta-signals and collision rules used in \ref{fig:update:delay} are listed in \RefFig{fig:def:delay}.
The signals \SigComputeMove and \SigComputeMoveOne (as well as \SigMinus and \SigMinusOne) are used to count down before disappearing after meeting both \SigL and \SigR (in whatever order).

\begin{figure}[hbt]
  \centering
  \newcommand{\DoListFwd}[1]{\csname Sig#1ComputeMove\endcsname,
    \csname Sig#1ComputeMoveOne\endcsname,
    \csname Sig#1Minus\endcsname,
    \csname Sig#1MinusOne\endcsname
  }\newcommand{\DoListBack}[1]{\csname Sig#1OneBack\endcsname,
    \csname Sig#1LBack\endcsname,
    \csname Sig#1RBack\endcsname,
    \csname Sig#1TestStart\endcsname
  }\begin{MSlist}
    &\DoListFwd{}, \DoListFwd{SR} &\SPEEDCOMPUTE\\
    &\DoListBack{}, \DoListBack{SR}, \SigSRPreComputeMove &-\SPEEDCOMPUTE\\
    &\DoListFwd{SL} &-\SPEEDCOMPUTE\\
    &\DoListBack{SL}, \SigSLPreComputeMove &\SPEEDCOMPUTE
  \end{MSlist}
  \\[.6em]
  \begin{CRlist}
    &\SigTree, \SigPreComputeMove &\SigTree, \SigComputeMove\\
    &\SigComputeMove, \SigOne &\SigOneBack, \SigOne, \SigComputeMove\\
    &\SigTree, \SigOneBack &\SigTree, \SigMinus\\
    &\SigTree, \SigTestStart &\SigTree, \SigTest
  \end{CRlist}
  \quad\begin{CRlist}
    $\forall \AGCmetaSignal \in \{\SigR, \SigL\}$&\SigComputeMove, \AGCmetaSignal &\AGCmetaSignal\IndexBack, \SigComputeMoveOne\\
    $\forall \AGCmetaSignal \in \{\SigR, \SigL\}$&\SigComputeMoveOne, \AGCmetaSignal &\AGCmetaSignal\IndexBack\\
    $\forall \AGCmetaSignal \in \{\SigR, \SigL\}$&\SigMinus, \AGCmetaSignal\IndexBack &\AGCmetaSignal, \SigMinusOne\\
    $\forall \AGCmetaSignal \in \{\SigR, \SigL\}$&\SigMinusOne, \AGCmetaSignal\IndexBack &\AGCmetaSignal, \SigTestStart\\
  \end{CRlist}
  \centerline{\footnotesize Collision rules with arrows are the same as without arrows.}

  \caption{Definitions for updating parameters after a \StepDelay.}
  \label{fig:def:delay}
\end{figure}

\begin{figure}[hbt]
  \centering
  \small\SetUnitlength{.45\textwidth}\hspace*{\fill}
  \subcaptionbox{\label{fig:delay:from delay}\StepDelay from \StepDelay}{\includegraphics[height=\unitlength]{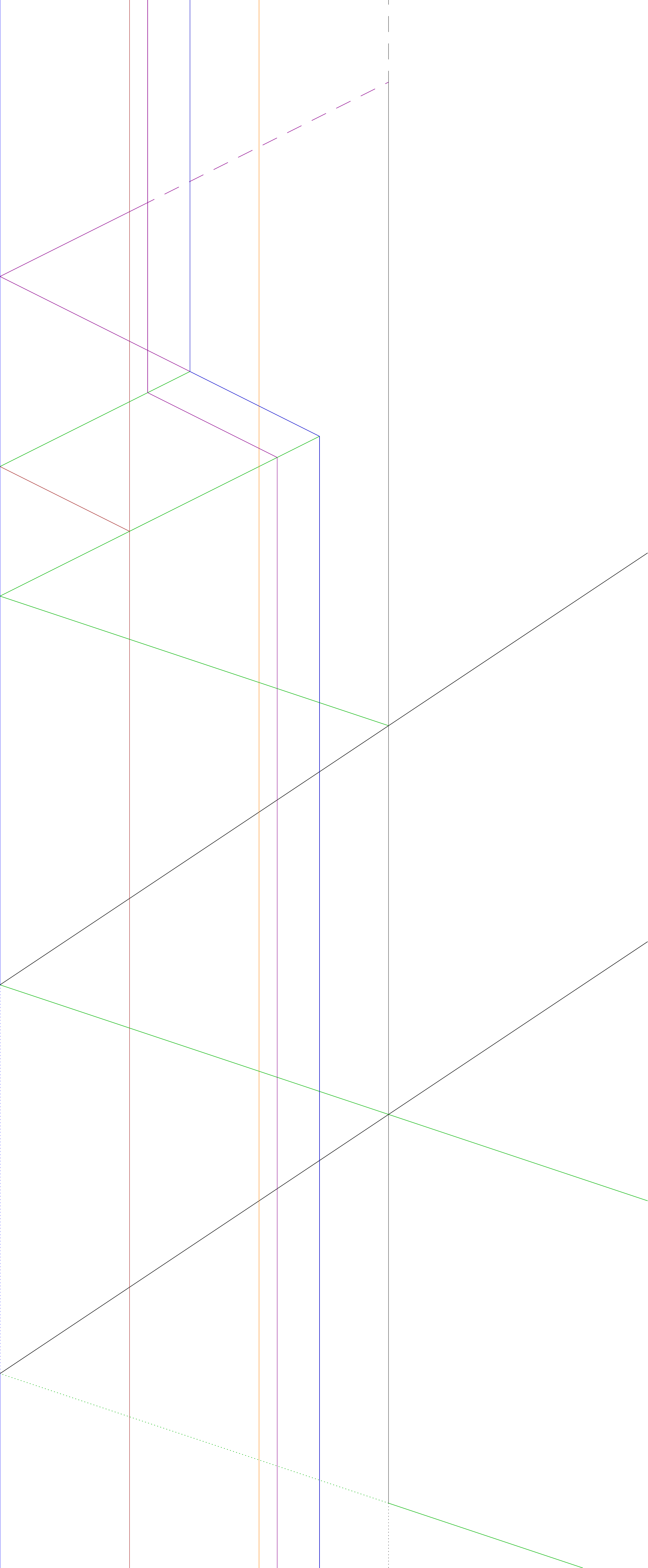}}\hfill\subcaptionbox{\label{fig:delay:from right}\StepDelay from right \StepSplit}{\includegraphics[height=\unitlength]{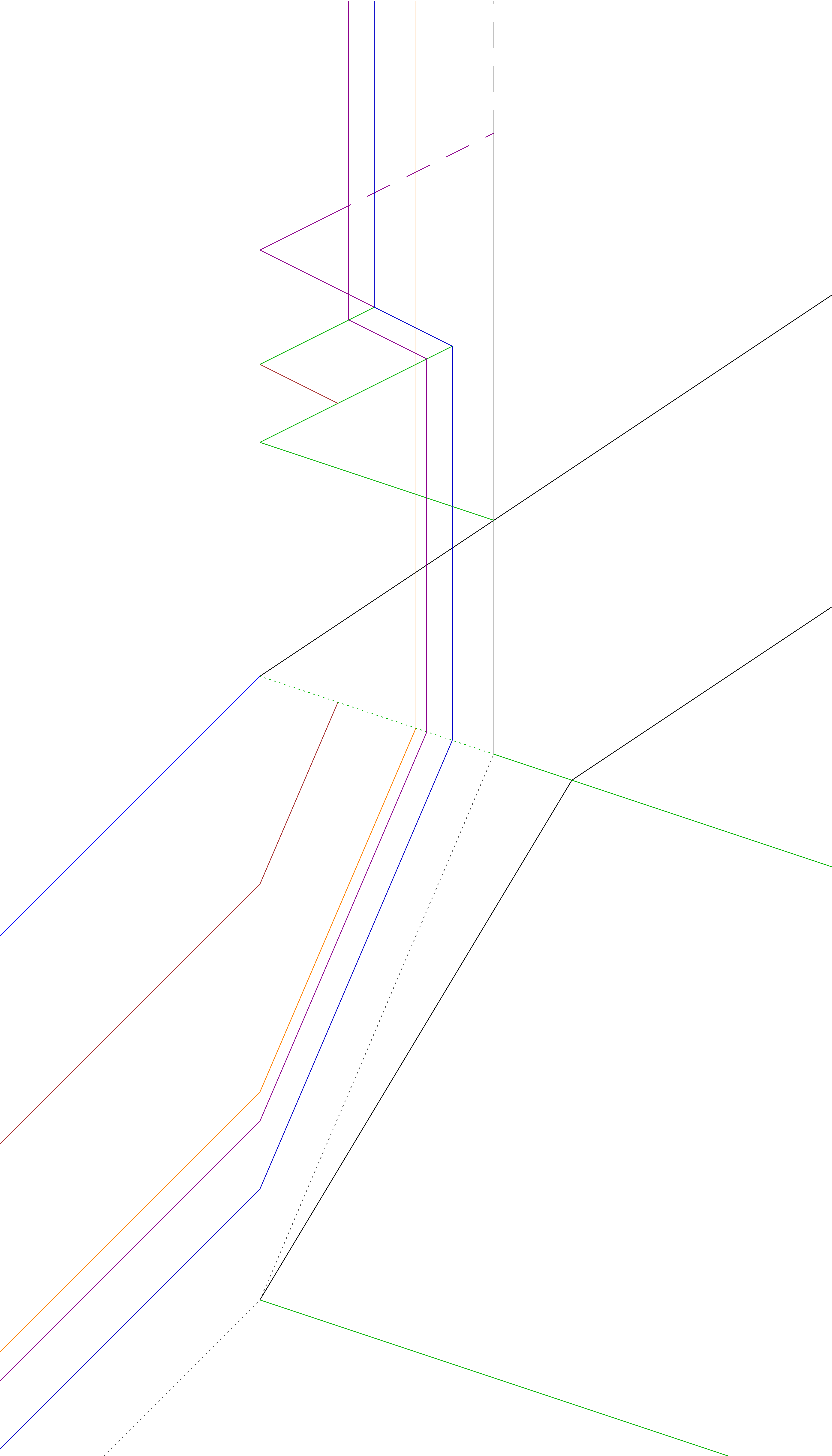}}\hfill\subcaptionbox{\label{fig:delay:from left}\StepDelay from left \StepSplit}{\includegraphics[height=\unitlength]{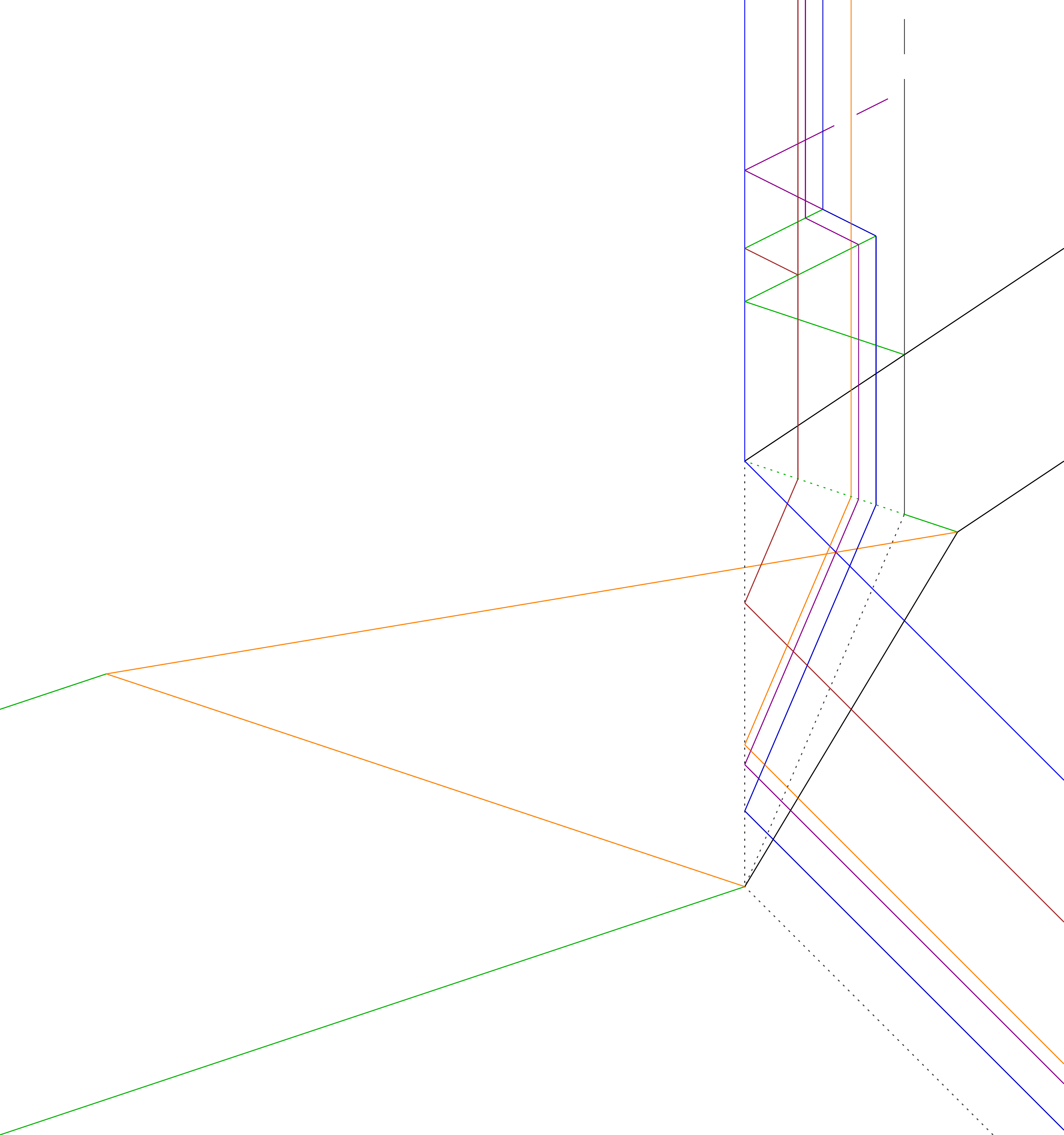}}\hspace*{\fill}
  \NTVS
  \caption{\StepDelay implementations.}
  \label{fig:delay}
\end{figure}

A full \StepDelay step, that is rerouting and parameter update, is performed as shown in \RefFig{fig:delay}.

\subsubsection{\StepSplit parameter update}

\begin{figure}[hbt]
  \centering\SetUnitlength{4.5em}\vspace*{\fill}\scalebox{.85}{\newcommand{\TOP}{4.8}\newcommand{\BOT}{-.7}\newcommand{\SPEEDlow}{3}\newcommand{\SPEEDhigh}{2}\begin{tikzpicture}[baseline={(\BOT,\BOT)}]
  \path (\BOT, \BOT) \CoorNode{treeInit} ;
  \path (0,0) \CoorNode{updtInit} ;
  \path (1 + \BOT,\BOT) \CoorNode{oneInit} ;
  \path (2 + \BOT,\BOT) \CoorNode{twoInit} ;
  \path (4.2 + \BOT, \BOT) \CoorNode{boundInit} ;
  \path (2.4 + \BOT, \BOT) \CoorNode{lInit} ;
  \path (1.3 + \BOT, \BOT) \CoorNode{rInit} ;
  \SetIntersectHeight[preInit]{updtInit}{-3}{\BOT}
  \SetIntersectHeight[treeTop]{updtInit}{1}{\TOP}
  \SetIntersectHeight[oneTop]{oneInit}{1}{\TOP}
  \SetIntersectHeight[twoTop]{twoInit}{1}{\TOP}
  \SetIntersectHeight[boundTop]{boundInit}{1}{\TOP}\SetIntersectLines[IntRlow]{updtInit}{\SPEEDlow}{rInit}{1}
  \SetIntersectLines[IntRhigh]{updtInit}{\SPEEDhigh}{rInit}{1}\SetIntersectLines[IntBackLow]{IntRlow}{-\SPEEDlow}{updtInit}{1}
  \SetIntersectLines[IntBackHigh]{IntRhigh}{-\SPEEDlow}{updtInit}{1}
  \SetIntersectLines[SetIntRlow]{IntBackLow}{\SPEEDlow}{rInit}{1}
  \SetIntersectLines[SetIntRhigh]{IntBackHigh}{\SPEEDlow}{SetIntRlow}{\SPEEDhigh}
  \SetIntersectLines[SetIntLlow]{IntBackLow}{\SPEEDlow}{lInit}{1}
  \SetIntersectLines[SetIntLhigh]{IntBackHigh}{\SPEEDlow}{SetIntLlow}{\SPEEDhigh}
  \SetIntersectHeight[lTop]{SetIntLhigh}{1}{\TOP}
  \SetIntersectHeight[rTop]{SetIntRhigh}{1}{\TOP}
  \SetIntersectLines[delayStart]{SetIntLhigh}{-\SPEEDlow}{updtInit}{1}
  \SetIntersectHeight[delayTop]{delayStart}{\SPEEDlow}{\TOP}
  \path[ray_style] (treeInit) -- (treeTop) -- (boundTop) -- (boundInit)
  -- cycle ;
  \DrawSigPreSplitUpdtRLUBelowLeft(updtInit)(preInit)
  \DrawSigSRTreeLUAbove(treeInit)(treeTop)
  \DrawSigSRBoundLUAbove(boundInit)(boundTop)
  \DrawSigSplitUpdtLowRLUBelow(updtInit)(IntRlow)
  \DrawSigSplitUpdtHighRLparam(updtInit)(IntRhigh)[sloped,above=-0.2em,pos=.4]
  \DrawSigSplitUpdtBackRLUAboveLeft(IntRhigh)(IntBackHigh)
  \DrawSigSplitUpdtBackRLUAboveLeft(IntRlow)(IntBackLow)
  \DrawSigSplitUpdtSetRLUAbove(IntBackLow)(SetIntRlow)
  \DrawSigSplitUpdtSetRLUAbove(IntBackHigh)(SetIntRhigh)
  \DrawSigSplitUpdtRSetRLUAbove(SetIntRlow)(SetIntRhigh)
  \DrawSigSRRLUBelowLeft(rInit)(SetIntRlow)\DrawSigSplitUpdtSetOneRLUBelow(SetIntRlow)(SetIntLlow)
  \DrawSigSplitUpdtSetOneRLUAbove(SetIntRhigh)(SetIntLhigh)
  \DrawSigSplitUpdtLSetRLUBelow(SetIntLlow)(SetIntLhigh)
  \DrawSigSRLLUBelowLeft(lInit)(SetIntLlow)\DrawSigSRLLUBelowRight(SetIntLhigh)(lTop)
  \DrawSigSRRLUBelowRight(SetIntRhigh)(rTop)
  \DrawSigSplitUpdtEndRLUAbove(SetIntLhigh)(delayStart)
  \DrawSigSRComputeMoveLUAbove(delayStart)(delayTop)
  \path[use as bounding box] (treeInit)  rectangle (boundTop) ;
\end{tikzpicture}}\vspace*{\fill}\hspace{-.35\textwidth}\vspace*{\fill}\includegraphics[width=.55\textwidth]{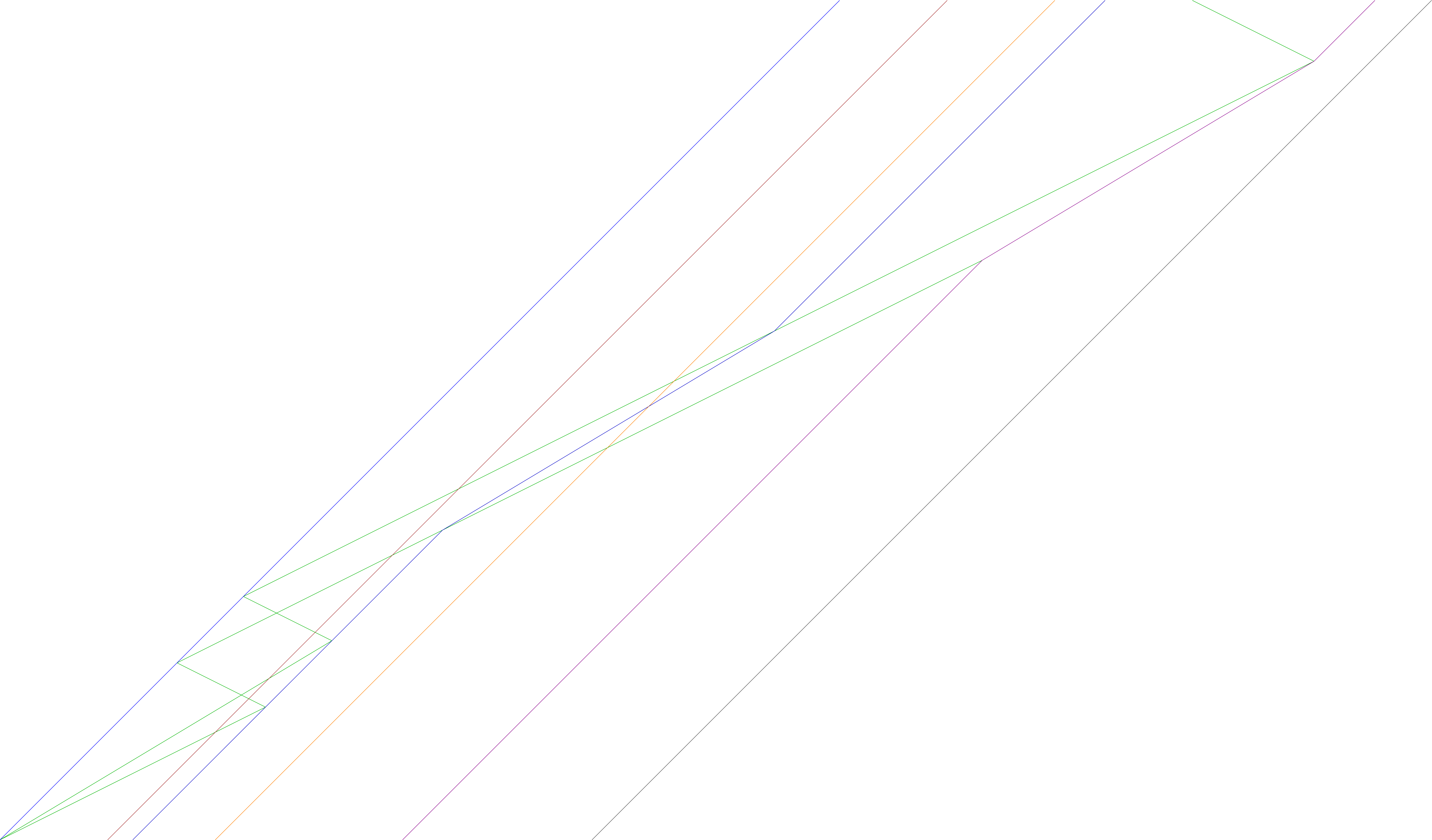}
  \caption{Updating \InputS and \InputT after a \StepSplit.}
  \label{fig:update:split}
\end{figure}

The updating on the right (resp. left) branch of a split is illustrated by \RefFig{fig:update:split} and done as follows: \InputT (resp. \InputS) is added to both \InputS and \InputT; then $1$ is removed from both \InputS and \InputT.
What happens on the right and on the left branch is symmetrical with the roles of \InputS and \InputT swapped; so that only what happens on a right branch is presented.

Adding \InputT to both \InputS and \InputT is done by moving \SigL and \SigR by the distance from \SigTree to \SigR (again using a parallelogram).
The computing signals bounce back to \SigSRTree after measuring that distance and before adding it so that the construction does not depend on the order of \SigL and \SigR.
The new meta-signals and collision rules at play are listed in \RefFig{fig:def:split:update}.

The computation is finished by removing $1$ and starting the test stage.
This is done exactly as in the parameter update occurring after a \StepDelay step.
The (dotted arrow) meta-signals and collision rules are already defined in \RefFig{fig:def:delay}.

\begin{figure}[hbt]
  \centering\begin{tabular}{c} 
    \SPEEDSPLITUP = \SPEEDSPLITUPvalue
  \end{tabular}
  \begin{MSlist}
    &\SigSplitUpdtLowR, \SigSplitUpdtSetR, \SigSplitUpdtSetOneR, \SigSplitUpdtBackL, \SigSplitUpdtEndL &\SPEEDCOMPUTE\\
    &\SigSplitUpdtLowL, \SigSplitUpdtSetL, \SigSplitUpdtSetOneL, \SigSplitUpdtBackR, \SigSplitUpdtEndR &-\SPEEDCOMPUTE\\
    &\SigSplitUpdtHighL, \SigSplitUpdtLSetL, \SigSplitUpdtRSetL &-\SPEEDSPLITUP\\
    &\SigSplitUpdtHighR, \SigSplitUpdtLSetR, \SigSplitUpdtRSetR &\SPEEDSPLITUP\\
  \end{MSlist}
  \medskip
  \begin{CRlist}
    $\dagger$ &\SigPreSplitUpdt, \AGCLeftRightSide{\SigTree} &\AGCLeftRightSide{\SigTree}, \SigSplitUpdtHigh, \SigSplitUpdtLow\\
    $\forall i \in\{\text{lo},\text{hi}\}$ &\SigSplitUpdtIR, \SigSRR &\SigSplitUpdtBackR, \SigSRR\\
    $\forall i \in\{\text{lo},\text{hi}\}$ &\SigSplitUpdtIL, \SigSLL &\SigSplitUpdtBackL, \SigSLL\\
    $\dagger$ &\AGCLeftRightSide{\SigTree}, \SigSplitUpdtBack &\AGCLeftRightSide{\SigTree}, \SigSplitUpdtSet\\
    $\dagger$ $\forall i \in \{\text{r}, \text{l}\}$ &\SigSplitUpdtSet, $\AGCLeftRightSide{i}$ &\SigSplitUpdtSetOne, \SigSplitUpdtISet\\
    $\dagger$ $\forall i \in \{\text{r}, \text{l}\}$ &\SigSplitUpdtSetOne, $\AGCLeftRightSide{i}$ &\SigSplitUpdtISet\\
    $\dagger$ $\forall i \in \{\text{r}, \text{l}\}$ &\SigSplitUpdtSetOne, \SigSplitUpdtISet & $\AGCLeftRightSide{i}$, \SigSplitUpdtEnd\\
    \end{CRlist}
  \\
  \footnotesize{$\dagger$ Each line starting with this symbol defines two rules, one with all right over arrows, one with all left.}
  \caption{Definitions for updating parameters after a \StepSplit.}
  \label{fig:def:split:update}
\end{figure}

\begin{figure}[hbt]
  \centering\tiny\subcaptionbox{\label{fig:split:delay}\StepSplit from \StepDelay}{\includegraphics[width=.6\textwidth]{019_X+Picture+cropped_slanted_split_delay.pdf}}\hfill\begin{tabular}[b]{c} 
    \subcaptionbox{\label{fig:split:from left}\StepSplit from left \StepSplit}{\includegraphics[width=.3\textwidth]{020_X+Picture+cropped_slanted_split_left.pdf}}
    \\
    \subcaptionbox{\label{fig:split:from right}\StepSplit from right \StepSplit}{\includegraphics[width=.3\textwidth]{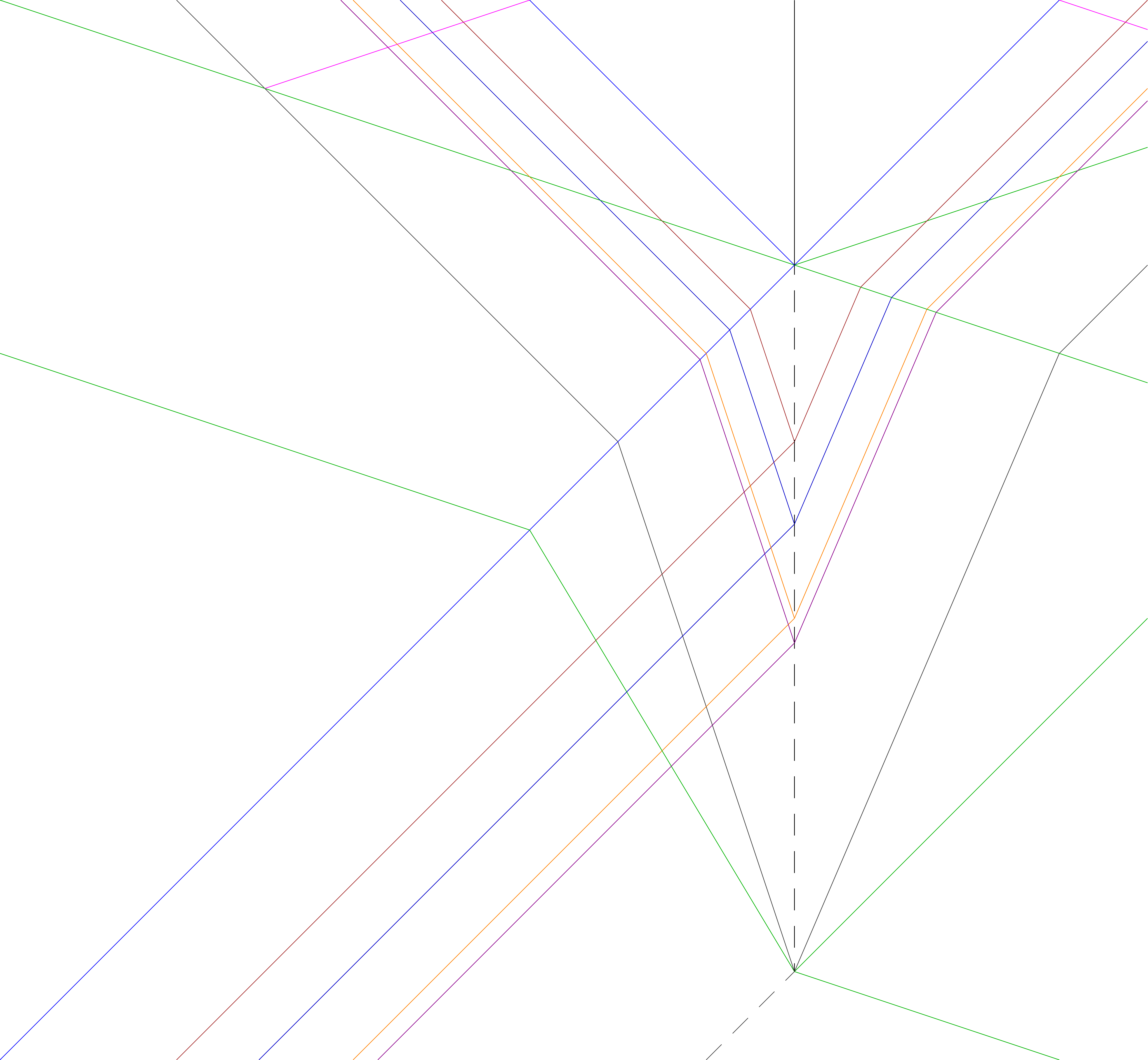}}
  \end{tabular}
  \NTVS
  \caption{\StepSplit implementations.}
  \label{fig:split}
\end{figure}

A full \StepSplit step, that is routing and parameter update is shown in \RefFig{fig:split}, with \RefFig{fig:split:from right} zooming on the routing part.

\subsection{Initial Configuration}
\label{subsec:initial configuration}

The initial configuration consists in a rightward tree macro-signal, built as per subsection \ref{subsec:Encoding}, as well as two bouncing signals (non slow and going right as well), separated by three times the width of the macro-signal.
A \SigBounceRTop signal is still inside the macro-signal, so as to prompt updating and testing.
The parameters inside the macro-signal at time 0 are chosen so as to become the targeted \InputS and \InputT after update.

\section{Conclusion}
\label{sec:conclusion}

\begin{figure}[hbt]
  \centering\includegraphics[width=.53\textwidth]{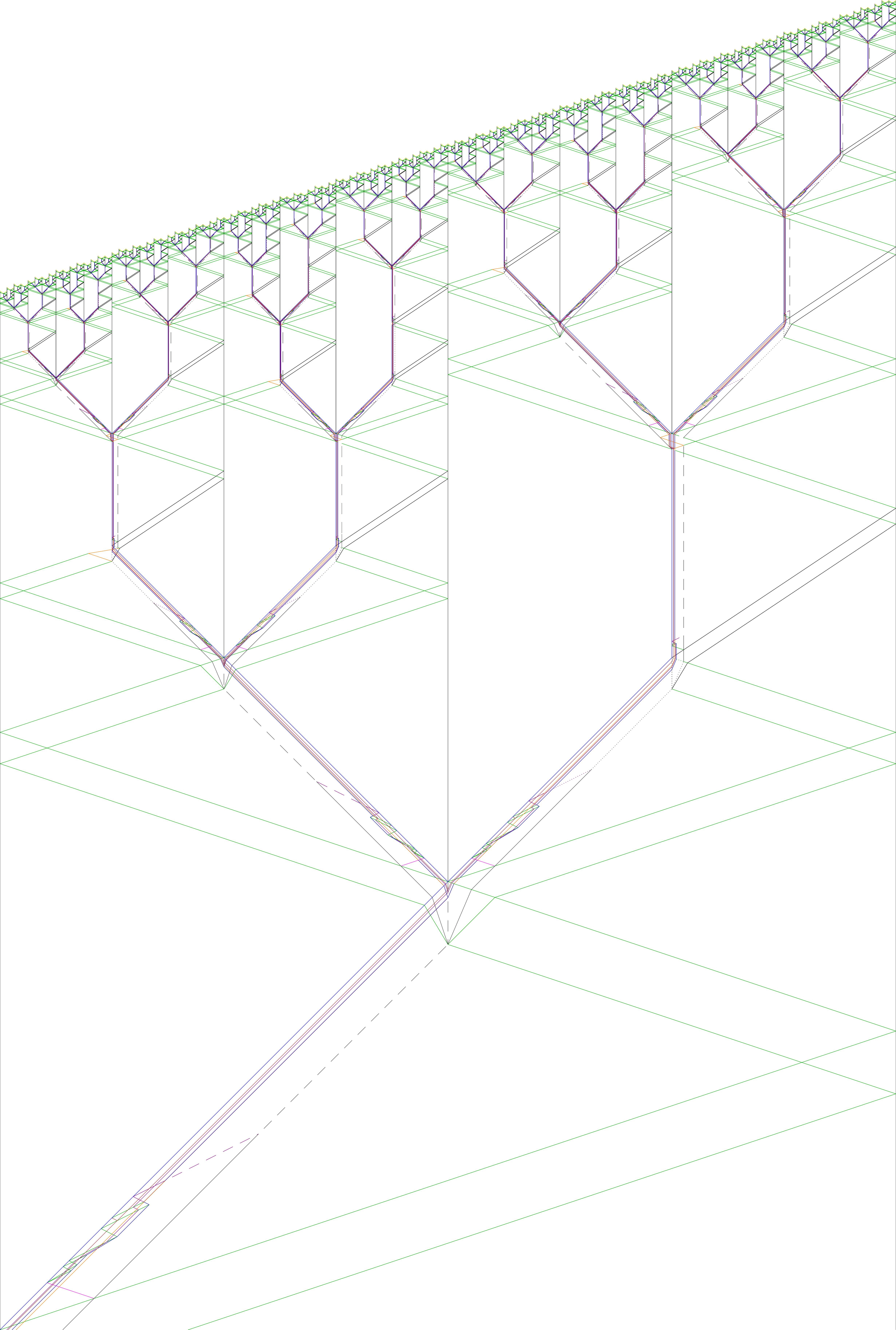}
  \NTVS
  \caption{The whole signal machine in action involving more than 74.000 collisions.}
  \label{fig:concl:example}
\end{figure}

We provide a recursive geometrical algorithm to draw an infinite tree accumulating on a parameterised slope together with its proof and its implementation as a signal machine.
An extended run on a signal machine can be seen in \RefFig{fig:concl:example}, with $\InputS=150/113$ and $\InputT=200/101$.

By comparison with cellular automata, the active signals (the general) start on an arbitrarily small area.
It corresponds to the single active cell at the start in cellular automata.

It takes several signals to code any slope with finitely many meta-signals.
It is possible to shrink the general to a single point and do similar construction to ours (for example by building bands of signals such as our).
There would, however, be only one attainable slope per general (assuming its position is fixed), and finitely many per signal machine (since point generals are one of finitely many meta-signals).

An important model that appears in the current construction is the augmented signal machine.
It allows to construct and prove at some level, leaving only ``technicalities'' to simulate with a usual signal machine (using macro-signals and macro-collisions).
These technicalities can be involving by themselves.

Our construction exhibits rational robustness: if the signal speed and initial positions are rational, so will be the targeted slope.
Further more the tree drawn is fractal (can be recursively defined) if and only if $\InputS$ and $\InputT$ are rational numbers.
Indeed, if say, $\InputS$ is irrational, then the path from the general to $(-1, \InputS)$ is made of a non-periodic sequence of \StepDelay and \StepSplit.
If $\InputS$ and $\InputT$ are rational, then one can show the set of reachable parameters is finite by observing the invariant: there is some positive integer $m$ (the least common multiple of the denominators of initial \InputS and \InputT) such that both $m.\InputS$ and $m.\InputT$ always belongs to $\NaturalSet$, as well as the fact that $\InputS$ and $\InputT$ are bounded.

\begin{figure}[hbt]
  \centering\includegraphics[width=.53\textwidth]{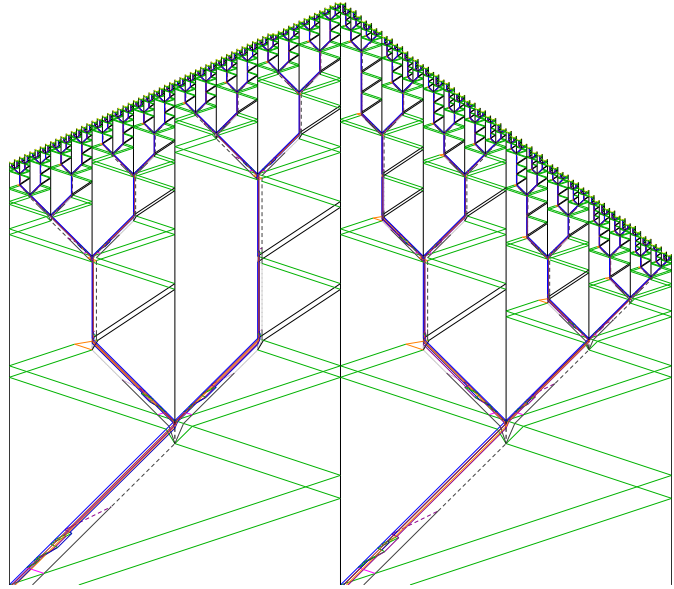}
  \NTVS
  \caption{Two different slopes.}
  \label{fig:concl:segments}
\end{figure}

This construction is a first step toward a study of possible accumulation sets of signal machines.
The next step is to look for curves, i.e. non piece-wise rectilinear.
We can already adapt the initial configuration (without modifying the machine!) to accumulate on a continuous piece-wise linear function, as shown in \RefFig{fig:concl:segments}.

\bibliographystyle{elsarticle-harv}

\bibliography{001_bib_jdl,002_bib_paper}

\end{document}